\documentclass[journal]{IEEEtran}

\usepackage{cite}
\usepackage{array}
\usepackage{color}
\usepackage{float}
\usepackage[cmex10]{amsmath}
\usepackage{nomencl}						
\usepackage[normalem]{ulem}			
\usepackage{diagbox}						
\usepackage{slashbox}						
\usepackage{colortbl}						
\usepackage{multirow}						
\usepackage{tabularx}
\usepackage{placeins}
\usepackage{algorithm}
\usepackage[noend]{algpseudocode}
\usepackage{siunitx}
\usepackage{cleveref}
\usepackage{multirow}

\usepackage{gensymb}

\usepackage{array, makecell}

\ifCLASSINFOpdf
  \usepackage[pdftex]{graphicx}
	\graphicspath{{./figure/}}
  \DeclareGraphicsExtensions{.pdf,.jpeg,.png}
\else
  \usepackage[dvips]{graphicx}
  \graphicspath{{./figure/}}
  \DeclareGraphicsExtensions{.eps}
\fi

\ifCLASSOPTIONcompsoc
  \usepackage[caption=false,font=normalsize,labelfont=sf,textfont=sf]{subfig}
\else
  \usepackage[caption=false,font=footnotesize]{subfig}
\fi

\newcommand{\figref}[1]{Fig.~\ref{#1}}
\newcommand{\tableref}[1]{Table~\ref{#1}}
\newcommand{\equref}[1]{(\ref{#1})}
\newcommand{\citeref}[1]{\cite{#1}}
\newcommand{\sectionref}[1]{Section~\ref{#1}}
\newcommand{\appendixref}[1]{Appendix~\ref{#1}}

\newcommand{\highlight}[1]{\textcolor{black}{#1}}

\begin{document}

\clearpage
\onecolumn
\thispagestyle{plain}
\twocolumn[
\begin{@twocolumnfalse}
	\begin{center}
	\vspace{4cm}
	\LARGE
	\textbf{Copyright Statements}
	\vspace{1cm}
	\end{center}
	\Large
	This work has been submitted to the IEEE for possible publication. Copyright may be transferred without notice, after which this version may no longer be accessible.
\end{@twocolumnfalse}]
\clearpage

\bstctlcite{IEEEexample:BSTcontrol}
\setcounter{page}{1}

\title{Revisiting Grid-Forming and Grid-Following Inverters: A Duality Theory}
\author{Yitong Li, \IEEEmembership{Member, IEEE}, Yunjie Gu, \IEEEmembership{Senior Member, IEEE}, Timothy C. Green, \IEEEmembership{Fellow, IEEE}}

\ifCLASSOPTIONpeerreview
	\maketitle 
\else
	\maketitle
\fi


\begin{abstract}
Power electronic converters for integrating renewable energy resources into power systems can be divided into grid-forming and grid-following inverters. They possess certain similarities, but several important differences, which means that the relationship between them is quite subtle and sometimes obscure. In this article, a new perspective based on \textit{duality} is proposed to create new insights. It successfully unifies the grid interfacing and synchronization characteristics of the two inverter types in a symmetric, elegant, and technology-neutral form. Analysis shows that the grid-forming and grid-following inverters are \textit{duals} of each other in several ways including a) synchronization controllers: frequency droop control and phase-locked loop (PLL); b) grid-interfacing characteristics: current-following voltage-forming and voltage-following current-forming; c) swing characteristics: current-angle swing and voltage-angle swing; d) inner-loop controllers: output impedance shaping and output admittance shaping; and e) grid strength compatibility: strong-grid instability and weak-grid instability. The swing equations are also derived in dual form, which reveal the dynamic interaction between the grid strength, the synchronization controllers, and the inner-loop controllers. Insights are generated into cases of poor stability in both small-signal and transient/large-signal. The theoretical analysis and simulation results are used to illustrate cases for simple single-inverter-infinite-bus systems and a multi-inverter power network.
\end{abstract}




\section{Introduction} \label{Section:Introduction}

The imperative to adopt low-carbon energy is bringing about connection of large volumes of renewable energy resources and other supporting resources (including wind turbines, solar arrays, battery storage and fuel cells) to power systems and these connections are via power electronic inverters \citeref{blaabjerg2006overview}. Compared with synchronous generators which have rotor motion dynamics and well-established standard models \citeref{kundur1994power}, the grid-interfacing and grid-synchronization characteristics of these inverter-based resources (IBRs) have dynamics set by control algorithms of which there are many variants \citeref{blaabjerg2006overview,wang2020grid,li2021impedance}. The central importance of the control algorithms of inverters brings flexibility to the system but with it come new control interaction and instability problems.

\begin{figure*}[t!]
\centering
\includegraphics[scale=1.07]{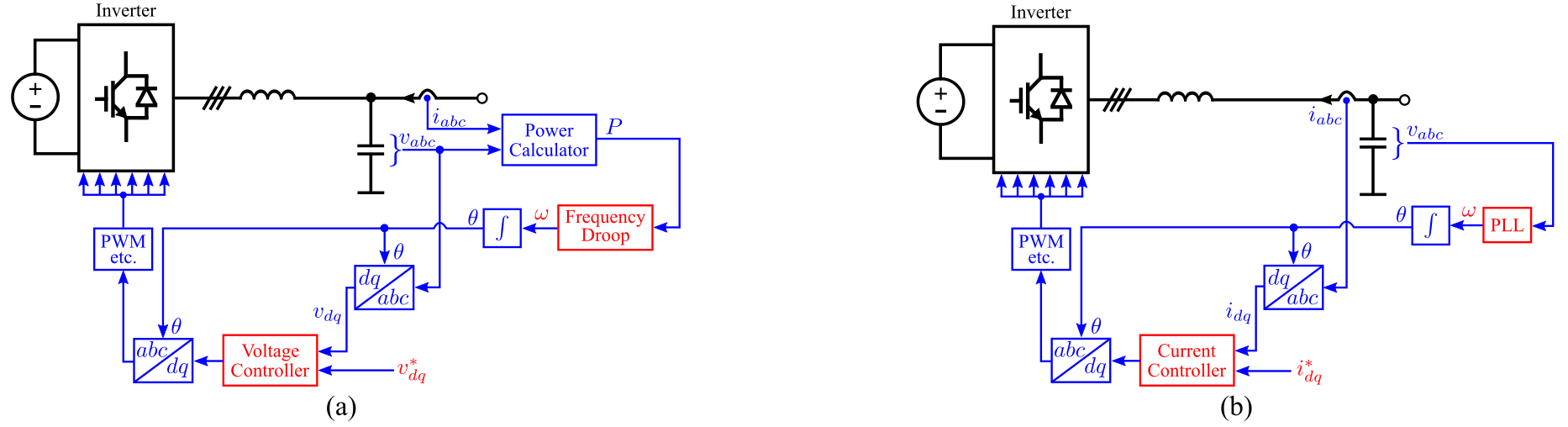}
\caption{Inverters and their control structures in synchronous $dq$ frame. (a) Grid-forming inverter with frequency droop control. (b) Grid-following inverter with PLL. (Note that the superscript $^*$ indicates the reference signal in this article.)}
\label{Fig:InverterControl}
\end{figure*}

Power electronic inverters used for IBRs are normally classified into two types, as shown in \figref{Fig:InverterControl} \footnote{In order to highlight the duality relationships and to facilitate comparison, the figures and tables in this paper are always arranged with the frequency droop grid-forming inverter case on the left and the PLL grid-following inverter case on the right.} : (a) grid-forming (GFM); and (b) grid-following (GFL). A grid-forming inverter controls the ac-side voltage and contributes to forming of a voltage-source grid. It synchronizes with the rest of the grid through frequency droop control (normally $P$-$\omega$ droop), which is similar to the control a synchronous generator\citeref{li2021impedance,darco2014equivalence,rosso2021grid}. In contrast, a grid-following inverter controls the ac-side current, and follows the phase angle of the existing grid voltage through a phase-locked loop (PLL) \citeref{chung2000phase,wang2020grid}. Grid-following inverters have already been widely used for integrating wind and solar energy into power grids due to their simple control structure, their mature PLL technology, and their feature of operating at a determined current (matching the maximum power point or dispatch point of the resource). However, the PLL has negative effects on the system stability especially when the grid is weak because of a large grid impedance \citeref{wen2015analysis,dong2015analysis}. This problem is likely to arise more often and become more challenging worldwide as more and more synchronous generators are replaced by IBRs. Consequently, attention has turned in recent years to grid-forming inverters because of their synchronous-generator-like characteristics and their capability of operating in a weak grid without stiff voltage sources or even of forming a stand-alone grid \citeref{li2021impedance,matevosyan2019grid,rosso2021grid}. 


\begin{figure*}[t!]
\centering
\includegraphics[scale=1.1]{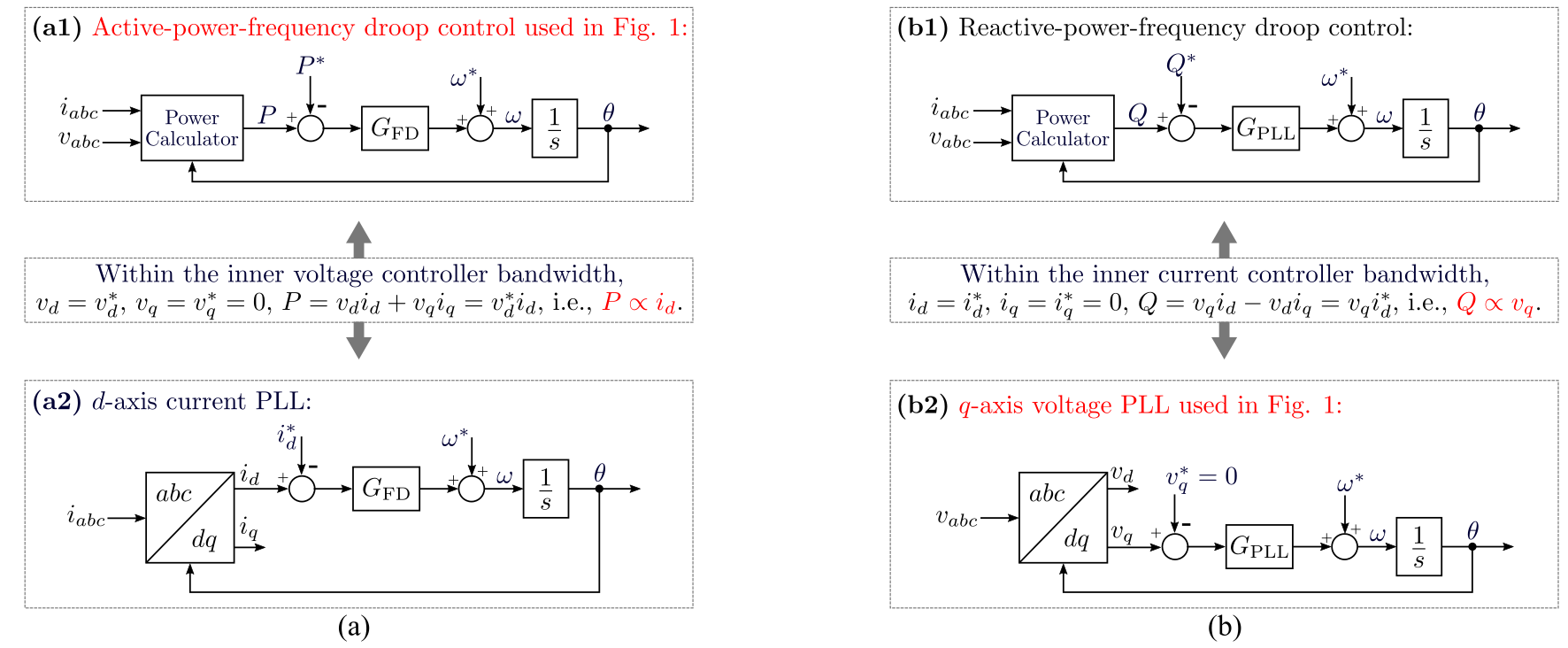}
\caption{Synchronization loops. (a) Frequency droop control for grid-forming inverter. (b) Synchronous-reference-frame PLL for grid-following inverter. (Notes: The current flowing into the inverter is assumed as the positive direction in this article, i.e., load convention.)}
\label{Fig:SynchronizationLoop}
\end{figure*}

As reviewed in \citeref{wang2020grid}, the characteristics of grid-forming and grid-following inverters are significantly different and they are seen as distinct  types of inverter with their own analysis tools and underpinning theory. However, intriguing similarities have also been reported in literature by researchers who have attempted to unify them. For example, \citeref{hu2019large,taul2019overview,zhong2016structural} observed the structural resemblance between the frequency droop control of a grid-forming inverter (or a synchronous generator) and the PLL of a grid-following inverter when the grid impedance is taken into account. Additionally, note that the PLL that was investigated in  \citeref{zhong2016structural} had been enhanced and is different from the widely-used fundamental PLL in \citeref{chung2000phase}. In \citeref{yuan2017modeling}, the small-signal power-frequency relation of a grid-following inverter was derived, but by merging the PLL with the dc-link voltage control to perform the analysis. These results are  inspiring indeed, but are not yet symmetrical or as elegant as they might be because of two straightforward obstacles \citeref{wang2020grid}: (a) a grid-forming inverter controls its ac-side voltage, but a grid-following inverter controls its ac-side current; (b) the frequency droop control establishes a relationship between frequency and active power, but a PLL establishes a relationship between frequency and grid voltage. These two differences have left the unification of two inverter types still an open question and has also obscured the relationship between.

In this article, a new perspective is developed that reveals a \textit{duality} in the relationship between the two inverter types in a technology-neutral form and provides an elegant framework to analyze them. The analysis will show that the frequency-droop control and the PLL of two inverter types are essentially \textit{duals} of each other. Further, the grid-forming and grid-following inverters can also be re-classified more precisely as \textit{current-following voltage-forming inverter} and \textit{voltage-following current-forming inverter}, respectively. By deriving and analyzing the swing characteristics, the synchronization mechanisms and grid-interfacing dynamics of the two types can also be re-defined as \textit{current-angle swing} and \textit{voltage-angle swing}, respectively. Based on the proposed duality theory, some questions and controversies that have concerned the power engineering community in recent times can be answered intuitively and convincingly. The questions include: a) How can the relationship between grid-forming and grid-following be understood? b) How do the controllers (e.g., droop control, PLL, and voltage or current loops) interact with each other? c) Is a grid-forming inverter also vulnerable to certain grid strength conditions just as a grid-following inverter is vulnerable to weak grids? d) Can a grid-following inverter also operate in islanding mode, like a grid-forming inverter? e) How to analyze the transient stability of grid-forming and grid-following inverters? Beyond answering these questions, the duality theory has the potential of inspiring new concepts and interpreting surprising phenomena, and providing fresh insights in power systems dominated by IBRs.

This article is organized as follows: \sectionref{Section:Duality} proposes a duality approach and elaborates it in various aspects. In \sectionref{Section:DualitySmallSignalStability} and \sectionref{Section:DualityTransientStability}, comprehensive analysis of swing characteristics reveals the duality of the synchronization stability of grid-forming and grid-following inverters in both small-signal and large-signal. Case studies are also investigated. Finally, \sectionref{Section:Conclusions} concludes the article.


\section{Duality of Grid-Forming and Grid-Following Inverters} \label{Section:Duality}

\begin{table*}[t!]
\renewcommand{\arraystretch}{1.3}
\newcommand{\tabincell}[2]{\begin{tabular}{@{}#1@{}}#2\end{tabular}}
\caption{Duality of Gird-Forming and Grid-Following Inverters}
\label{Table:DualitySummary}
\centering
\begin{tabular}{|c|c||l|l|}
\hline
\multicolumn{2}{|c||}{} & ~~~\tabincell{c}{~\\ \textbf{Frequency Droop Grid-Forming Inverter}\\~} & ~~~~~~~~~~~~~\textbf{PLL Grid-Following Inverter} 
\\
\hline
\hline
\multicolumn{2}{|c||}{\tabincell{c}{Synchronization\\Control}} & \tabincell{l}{$P$-$\omega$ droop control with droop gain $G_\text{FD}$.\\($i_d$-$\omega$ droop control with droop gain $G_\text{FD}$.)\\($i_d$-PLL with phase-locking controller $G_\text{FD}$.)} & \tabincell{l}{$v_q$-PLL with phase-locking controller $G_\text{PLL}$.\\($v_q$-$\omega$ droop control with droop gain $G_\text{PLL}$.)\\($Q$-$\omega$ droop control with droop gain $G_\text{PLL}$.)} 
\\
\hline
\multicolumn{2}{|c||}{\tabincell{c}{Grid-Interfacing\\Characteristics}} & Grid current-following voltage-forming. & Grid voltage-following current-forming.
\\
\hline
\multicolumn{2}{|c||}{\tabincell{c}{Swing\\Characteristics}} & \tabincell{l}{$I$-$\theta$ swing or $P$-$\theta$ swing.} & \tabincell{l}{$V$-$\theta$ swing or $Q$-$\theta$ swing.}
\\
\hline
\multicolumn{2}{|c||}{\tabincell{c}{Extreme\\Operation}} & \tabincell{l}{Stable when open-circuit with $Z_g \rightarrow \infty$, or\\
~~~~connected to an ideal current source.\\Unstable when short-circuit with $Z_g=0$, or\\~~~~connected to an ideal voltage source.} & \tabincell{l}{Stable when short-circuit with $Y_g \rightarrow \infty$, or\\~~~~connected to an ideal voltage source.\\Unstable when open-circuit with $Y_g = 0$, or\\~~~~connected to an ideal current source.}
\\
\hline
\multirow{4}{*}{\rotatebox[origin=c]{90}{\parbox[c]{2cm}{\centering Interaction}}} & \tabincell{c}{Grid Strength} & \tabincell{l}{Unstable when grid is strong with small $Z_g$.\\(Weak-grid-current-strength instability;\\Strong-grid-voltage-strength instability.)} & \tabincell{l}{Unstable when grid is weak with small $Y_g$.\\(Weak-grid-voltage-strength instability;\\Strong-grid-current-strength instability.)}
\\
\cline{2-4}
& \tabincell{c}{Synchronization\\Controller} & \tabincell{l}{Unstable when $G_\text{FD}$ (e.g., droop gain $m$) is large,\\~~~~i.e., $Z_\text{FD}$ is large.} & \tabincell{l}{Unstable when $G_\text{PLL}$ (e.g., PLL bandwidth) is large,\\~~~~i.e., $Y_\text{PLL}$ is large.} 
\\
\cline{2-4}
& \tabincell{c}{Inner Loop\\Controller} & \tabincell{l}{Unstable when voltage control bandwidth is low,\\~~~~i.e., $Z_c$ is large.} & \tabincell{l}{Unstable when current control bandwidth is low,\\~~~~i.e., $Y_c$ is large.}.
\\
\hline
\end{tabular}
\end{table*}

\subsection{Duality of Synchronization Loops} \label{Section:DualitySynchronizationLoop}

The synchronization functions of inverters are highlighted in red in \figref{Fig:InverterControl}, namely a frequency droop block for the grid-forming inverter and a PLL for the grid-following inverter. The control block diagram format for these two synchronization methods is illustrated in \figref{Fig:SynchronizationLoop}.

Examining first the droop control of a grid-forming inverter in (a1) in \figref{Fig:SynchronizationLoop}(a), we recall that $P=v_di_d + v_qi_q$, and note that, if  $v_d$ is constant and $v_q$ is zero, $P$-$\omega$ droop in (a1) is equivalent to $i_d$-$\omega$ droop in (a2). This is valid within the voltage-loop bandwidth of the grid-forming inverter and with a zero reference for $q$-axis voltage. The droop controller $G_\text{FD}$ normally has the form $m\frac{1}{1+s T_f}$ where $m$ is the droop gain and $\frac{1}{1+s T_f}$ is a low-pass filter (LPF) which attenuates high-frequency noise. As will be discussed in \sectionref{Section:DualitySmallSignalStability} and as noted in \citeref{li2021impedance,darco2014equivalence}, this LPF also provides virtual inertia. For the grid-following inverter, (b2) in \figref{Fig:SynchronizationLoop}(b) shows the widely-used synchronous-reference-frame PLL. The PLL measures the voltage phase angle by controlling $v_q$ to zero through a proportional-integral (PI) controller $G_\text{PLL} = k_{p,\text{PLL}} + \frac{k_{i,\text{PLL}}}{s}$. Recalling that reactive power is given by $Q=v_qi_d - v_di_q$ we note that $v_q$ is also proportional to $Q$  when $i_d$ is constant and $i_q$ is zero. This is valid within the current loop bandwidth (with a reference of zero for $q$-axis current). In other words, the $v_q$-PLL in (b2) is equivalent to $Q$-PLL in (b1).

By comparing \figref{Fig:SynchronizationLoop}(a) and (b) carefully, the duality can be identified. Specifically, a power-frequency droop is also a $i_d$-$\omega$ droop that can be regarded as a PLL that tracks $i_d$. This $i_d$-PLL has a phase-locking controller $G_\text{FD}$ that is a proportional (droop) gain rather than a PI controller and hence steady-state tracking error is expected. As a dual, a PLL that tracks $v_q$ can also be regarded as a $v_q$-$\omega$ droop control. Its controller $G_\text{PLL}$ is a PI droop gain without steady-state droop errors. In summary, and when putting aside the difference between $G_\text{FD}$ and $G_\text{PLL}$, the two synchronization methods are perfectly duals of each other: one tracks (or droops) $i_d$ or $P$ whereas the other tracks (or droops) $v_q$ or $Q$, as summarized in \tableref{Table:DualitySummary}. This duality view of  the two synchronization methods drives rethinking the grid-forming and grid-following concepts that is expounded in the following subsections.



\subsection{Duality of Grid-Forming and Grid-Following} \label{Section:DualityGridFormingFollowing}

A very important point that is so common and often overlooked is recalled here: power grids nowadays are voltage sources by default and the connections of sources and loads are made in parallel. This choice has been driven by efficiency and power quality considerations \citeref{brameller1980electric}. In other words, the grid-forming and grid-following are more precisely voltage-forming and voltage-following. Additionally, the $i_d$-$\omega$ droop ($i_d$-PLL) of the grid-forming inverter implies that it is current-following; and the current control of the grid-following inverter implies that it is current-forming. Hence, a frequency droop grid-forming inverter can be re-defined more precisely as a \textit{current-following, voltage-forming inverter} and, as a dual, a PLL grid-following inverter can also be re-defined more precisely as a \textit{voltage-following, current-forming inverter}. In other words, the grid-interfacing features of two inverters are also perfectly duals in a technology-neutral form, as summarized in \tableref{Table:DualitySummary}.


\subsection{Duality of Swing Characteristics} \label{Section:DualitySwing}

It has been established that the frequency droop grid-forming inverter operates similarly (or in certain cases equivalently) to a virtual synchronous generator and synchronizes to a grid system through swing dynamics \citeref{li2021impedance,darco2014equivalence}. By contrast, the swing of a PLL grid-following inverter is still ill-defined and under-researched \citeref{hu2019large,yuan2017modeling}. The swing characteristics of both inverters are derived in the next passage and further aspects of their duality are revealed.

\begin{figure*}[t!]
\centering
\includegraphics[scale=1]{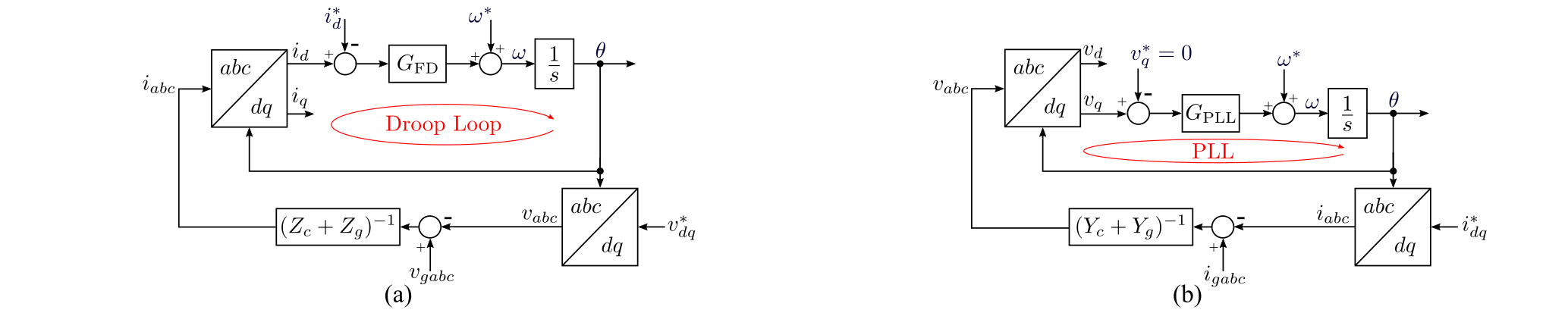}
\caption{Synchronization loops for single-inverter-infinite-bus systems. (a) Frequency droop grid-forming inverter. (b) PLL grid-following inverter.}
\label{Fig:SynchronizationFullLoop}
\end{figure*}

\begin{figure*}[t!]
\centering
\includegraphics[scale=1]{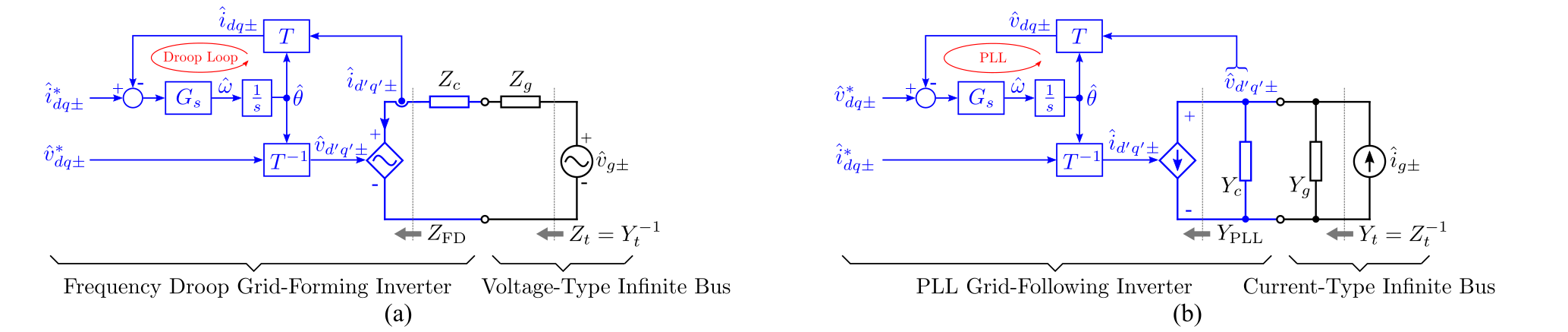}
\caption{Modeling single-inverter-infinite-bus systems in $dq\pm$ frame. (a) Frequency droop grid-forming inverter. (b) PLL grid-following inverter. (Notes: These models are small-signal, and in complex $dq$ frame. $G_s$ is the synchronization controller. $T$ is the transformation matrix from $d^\prime q^\prime \pm$ frame to $dq\pm$ frame.)}
\label{Fig:SingleInverterInfiniteBus}
\end{figure*}

\begin{figure}[t!]
\centering
\includegraphics[scale=1]{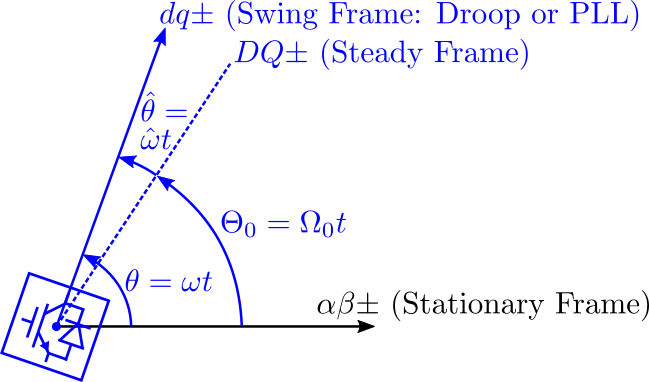}
\caption{Inverter frame illustration. (Notes: $\omega$ and $\theta$ are the large-signal angular frequency and phase angle respectively, with small-signal $\hat{\omega}$ and $\hat{\theta}$ and steady-state operating points $\Omega_0$ and $\Theta_0$.)}
\label{Fig:FrameDynamics}
\end{figure}

\begin{figure*}[t!]
\centering
\includegraphics[scale=1]{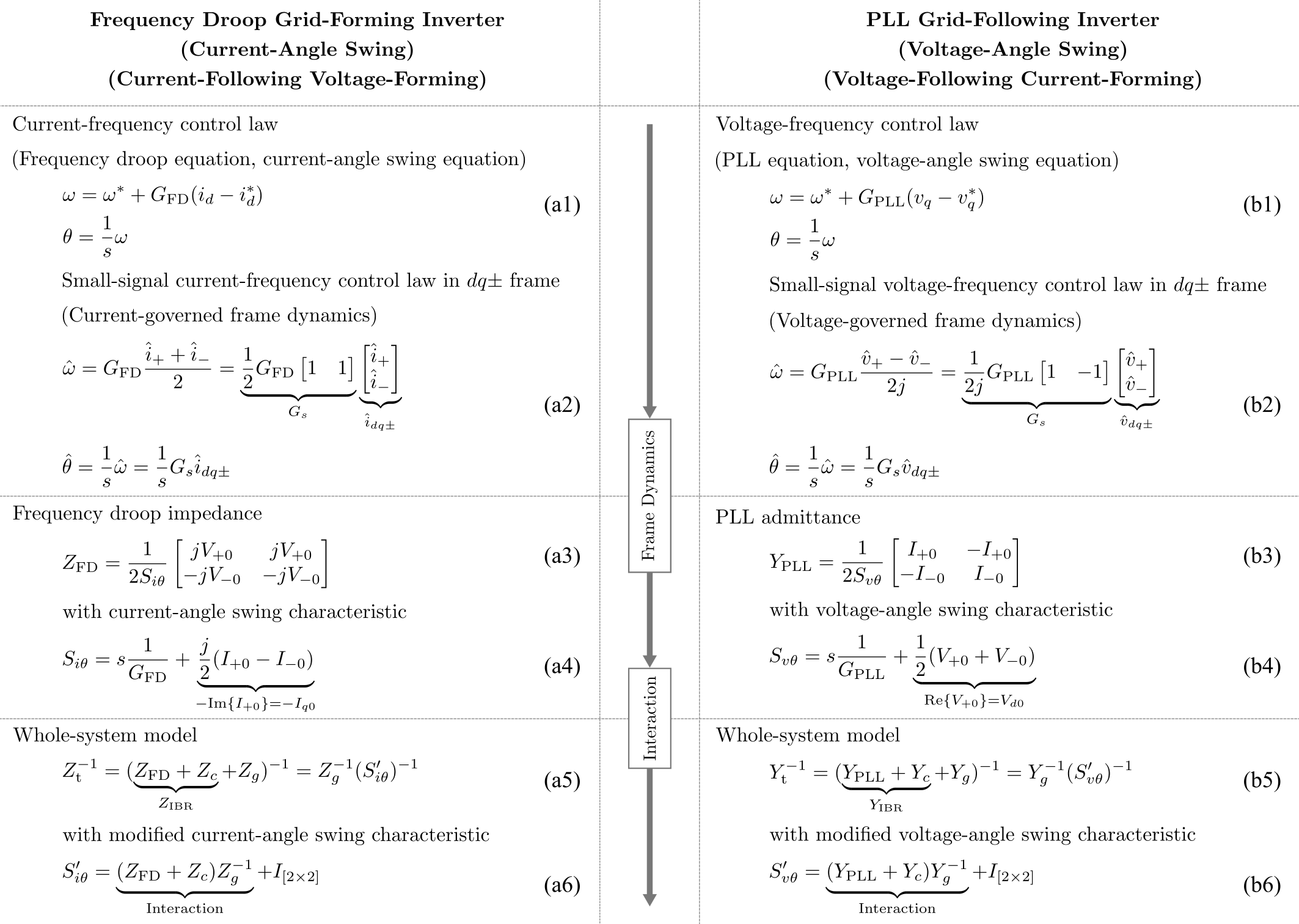}
\caption{Characteristic equations. (a) Frequency droop grid-forming inverter. (b) PLL grid-following inverter. (Notes: $s$ is the Laplace operator. $I_{[2\times 2]}$ is the $2 \times 2$ identity matrix. The linearization law includes $\omega = \hat{\omega} + \Omega_0$, $\theta = \hat{\theta}+\Theta_0$, $v = \hat{v} + V_0$, and $i = \hat{i}+I_0$.)}
\label{Fig:ModelDerivation}
\end{figure*}


\subsubsection{Frequency droop grid-forming inverter} The single-machine-infinite-bus system is a useful simple case to investigate the swing dynamics \citeref{kundur1994power,li2021impedance}. \figref{Fig:SynchronizationFullLoop}(a) gives a model of the synchronization process similar to \figref{Fig:SynchronizationLoop}(a) but with an additional feedback path considering grid impedance $Z_g$ and inverter impedance $Z_c$. The representation of the inverter as a controlled voltage source reflects the current-following voltage-forming feature and the synchronization loop. Impedance $Z_c$ incorporates the dynamics of the ac filter and of the inner voltage loop of the converter \citeref{li2021impedance}.
\figref{Fig:SingleInverterInfiniteBus}(a) is a re-casting of the model in small-signal form (for which variables are designated with $\hat{ }~$) and in the complex vector $dq$ frame (variables designated with $dq\pm$) \citeref{li2021impedance,li2020interpretating,harnefors2007modeling} \footnote{We use complex $dq$ frame ($dq\pm$ frame) for the small-signal analysis in this article, which represents $dq$ frame variables as space vectors and has more concise expression. The fundamentals of $dq\pm$ frame are summarized in \appendixref{Appendix:ComplexFrame}.}. 

The physical plant, the grid and inverter are represented in the steady frame ($d^\prime q^\prime\pm$) rotating  at constant speed $\Omega_0$ whereas the synchronization loop and voltage reference are represented in the local swing frame ($dq\pm$) rotating at frequency $\omega$ given by the frequency droop control. The relationship between the steady and swing frames is illustrated in \figref{Fig:FrameDynamics} \citeref{gu2021impedance}. \footnote{The swing frame and steady frames are also known as controller frame and system frame respectively in \citeref{wen2015analysis}, and also known as mechanical frame and electrical frame respectively in \citeref{li2021mapping}.} The frame transformation $[T]$ between $dq\pm$ and $d^\prime q^\prime \pm$ reflects the synchronization dynamics \citeref{gu2021impedance,li2021mapping}.

\figref{Fig:ModelDerivation} shows in the left-hand column the key steps in modelling the \textit{current-angle swing} dynamics of a grid-forming inverter. \highlight{(a1)} is the frequency droop equation and \highlight{(a2)} is linearized from using $dq\pm$ variables.  These equations show the swing relationship between the current and synchronization angle.

Later, in \sectionref{Section:MultiInverter}, the impedance method will be used to facilitate the analysis of multi-inverter cases and so the swing equation is developed in impedance form using the frame-dynamics-embedding method \citeref{gu2021impedance,li2021mapping}. Starting with \highlight{(a2)} and \figref{Fig:SingleInverterInfiniteBus}(a), we can obtain the frequency droop virtual impedance $Z_\text{FD}$ of \highlight{(a3)}, where $S_{i\theta}$ is the current-angle swing characteristic equation of \highlight{(a4)}. A whole-system model that considers both the inverter and the infinite bus is then created in \highlight{(a5)} which uses the modified swing characteristic equation $S_{i\theta}^\prime$ of \highlight{(a6)}. The modified swing characteristic captures the interaction between the synchronization loop, the inner voltage loop, and external grid impedance. The representation of the voltage-loop $Z_c$ can be found in literature  \citeref{li2021impedance}. Further analysis of $S_{i\theta}$ and $S_{i\theta}^\prime$ will be given in \sectionref{Section:DualitySmallSignalStability}.

\begin{figure}[t!]
\centering
\includegraphics[scale=1]{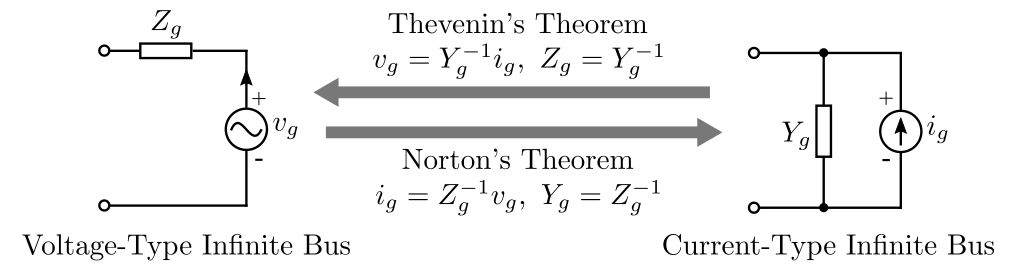}
\caption{Transformation between voltage-type and current-type infinite buses.}
\label{Fig:InfiniteBus}
\end{figure}

\subsubsection{PLL grid-following inverter} The synchronization loop including grid feedback is shown in \figref{Fig:SynchronizationFullLoop}(b) and its small-signal model in complex $dq$ form is shown in \figref{Fig:SingleInverterInfiniteBus}(b). The grid-following inverter is repented by a controlled current source which reflects the voltage-following current-forming feature. The admittance $Y_c$ represents the dynamics of the ac filter and the inner current loop \citeref{li2021impedance}. To highlight the duality between grid-forming and grid-following inverters, the infinite bus is represented by a current source with parallel admittance $Y_g$ here.  This is equivalent to a voltage-type infinite bus according to the Thevenin's and Norton's theorems, as illustrated in \figref{Fig:InfiniteBus}.

The right-hand column in \figref{Fig:ModelDerivation} sets out the swing equation of a PLL grid-following inverter. \highlight{(b1)} is the PLL swing equation and the linearized form is given in \highlight{(b2)}. The duality between inverter types is seen in the grid-forming inverter having current-angle swing characteristics whereas the grid-following inverter has \textit{voltage-angle swing} characteristics.

Following similar steps as for the frequency droop grid-forming inverter, the frame-dynamics-embedding method in \citeref{gu2021impedance,li2021mapping} is used to form a virtual admittance $Y_\text{PLL}$ representation of the PLL of \highlight{(b3)} with the voltage-angle swing characteristic equation $S_{v\theta}$ in \highlight{(b4)}. The whole-system model in \highlight{(b5)} can then be obtained, in which the current-loop virtual admittance $Y_c$ can be found in \citeref{li2021impedance}. $S_{v\theta}^\prime$ is the swing characteristic defined in \highlight{(b6)} which incorporates gird interaction. Further analysis of $S_{v\theta}$ and $S_{v\theta}^\prime$ will also be given in \sectionref{Section:DualitySmallSignalStability}, alongside that for the grid-forming inverter.

Summarising \figref{Fig:ModelDerivation}, the grid-forming inverter has current-angle swing characteristics (equivalent to $P$-angle swing within its voltage loop bandwidth when $v_q$ is constant and $v_q$ is zero) whereas, as a dual, the grid-following inverter has voltage-angle swing characteristics (equivalent to $Q$-angle swing when $i_d$ is constant and $i_q$ is zero). 

\subsection{Summary of Duality}

In this section, grid-forming and grid-following inverters have been discussed in terms of synchronization loops, the forming or following of voltage and current, and the swing dynamics. The discussion has applied the same perspective to each type and shown that in many regards, the two types have duality characteristics, as summarized in \tableref{Table:DualitySummary}. In next section, the analysis will be extended to stability of the synchronization to explore the duality further.



\section{Duality of Small-Signal Stability} \label{Section:DualitySmallSignalStability}

In this section, the swing characteristics in \figref{Fig:ModelDerivation} will be further analyzed, to explore the duality from the perspective of stability. Case-study networks will be used to illustrate the points. The parameters used for obtaining theoretical and simulation results in this section are summarized in \appendixref{Section:Paramters}.


\subsection{Synchronization Loops with Ideal Sources: $S_{i\theta}$ and $S_{v\theta}$} \label{Section:SingleInverterIdealSource}

The synchronization loops illustrated in \figref{Fig:SynchronizationLoop}(a) and (b) are investigated first for the simple case where the current vector detected by the frequency droop and the voltage vector detected by the PLL are steady vectors rotating with a constant angular frequency. In other words, the grid-forming inverter is required to synchronize to an ideal current source and the grid-following inverter to an ideal voltage source, without inner-loop dynamics and grid impedance. The synchronization dynamics are as described by $S_{i\theta}$ and $S_{v\theta}$ in \highlight{(a4) and (b4)} in \figref{Fig:ModelDerivation}. It is worth remarking that $S_{i\theta}$ can also be derived directly by linearizing the frequency droop equation noting that $\hat{i}_d = I_{q0}\hat{\theta}$; and $S_{v\theta}$ can be derived directly by linearizing the PLL equation noting that $\hat{v}_q = -V_{d0}\hat{\theta}$.

Taking $G_\text{FD}$ as a droop gain with an LPF ($m\frac{1}{1+T_f s}$) and $G_\text{PLL}$ as a PI controller ($k_{p,\text{PLL}} + \frac{k_{i,\text{PLL}}}{s}$), $S_{i\theta}$ and $S_{v\theta}$ can be written as
\begin{equation} \label{Equ:SwingGFM}
S_{i\theta} = s^2 \underbrace{\frac{T_f}{m}}_{J} + s \underbrace{\frac{1}{m}}_{K_D} + \underbrace{(-I_{q0})}_{K_S}
\end{equation}
and
\begin{equation} \label{Equ:SwingGFL}
S_{v\theta} = \frac{1}{1+s\frac{k_{p,\text{PLL}}}{k_{i,\text{PLL}}}}\bigg( s^2 \underbrace{\frac{1}{k_{i,\text{PLL}}}}_J + s\underbrace{\frac{k_{p,\text{PLL}}}{k_{i,\text{PLL}}}V_{d0}}_{K_D} + \underbrace{V_{d0}}_{K_S} \bigg)
\end{equation} 
The forms of $S_{i\theta}$ and $S_{v\theta}$ are not perfectly duals here because of the difference between controllers $G_\text{FD}$ and $G_\text{PLL}$, but, notwithstanding that, both $S_{i\theta}$ and $S_{v\theta}$ have characteristics similarly to a virtual synchronous generator with inertia $J$, damping torque coefficient $K_D$, and synchronizing torque coefficient $K_S$ as identified in the equations. In particular, the time-constant of the LPF in $G_\text{FD}$ and the integral gain of the PI controller in $G_\text{PLL}$ contribute the virtual inertia $J$. 


Unlike this simple case, the current and voltage vectors measured by frequency droop and PLL are not constant in practice, and further analysis of $S_{i\theta}$ and $S_{v\theta}$ is needed next.

\subsection{Single-Inverter-Infinite-Bus Systems: $S_{i\theta}^\prime$ and $S_{v\theta}^\prime$} \label{Section:SingleInverterInfiniteBus}

For single-inverter-infinite-bus systems (\figref{Fig:SynchronizationFullLoop} and \figref{Fig:SingleInverterInfiniteBus}) the current and voltage at the connection point to the inverters synchronize are not constant, and instead are subject to interactions between the grid impedance/admittance ($Z_g$ and $Y_g$), the synchronization controllers ($Z_\text{FD}$ and $Y_\text{PLL}$), and the inner-loop controllers ($Z_c$ and $Y_c$). The swing characteristics $S_{i\theta}$ and $S_{v\theta}$ are modified to $S_{i\theta}^\prime$ and $S_{v\theta}^\prime$ through the introduction of interaction terms $(Z_\text{FD}+Z_c)Z_g^{-1}$ and $(Y_\text{PLL}+Y_c)Y_g^{-1}$ respectively.

\begin{figure}[t!]
\centering
	\subfloat[]{\includegraphics[scale = 0.75]{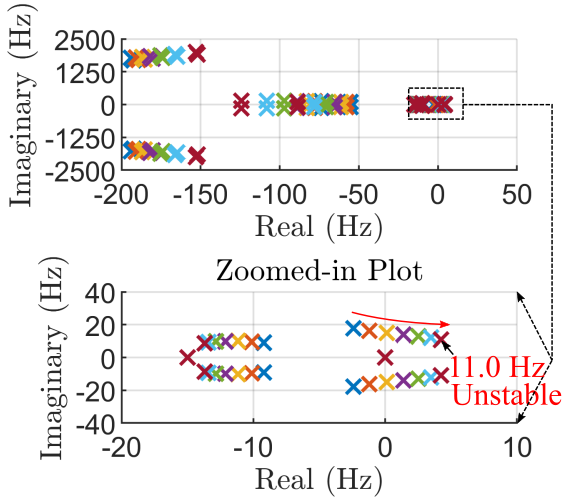}}
	\subfloat[]{\includegraphics[scale = 0.75]{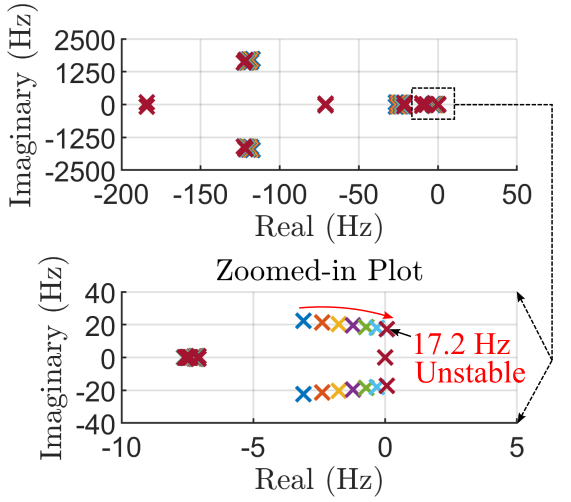}}
	
    \subfloat[]{\includegraphics[scale = 0.72]{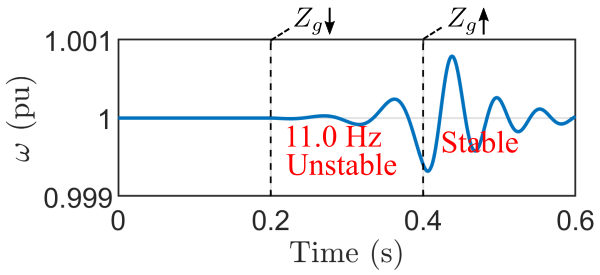}}
	\subfloat[]{\includegraphics[scale = 0.72]{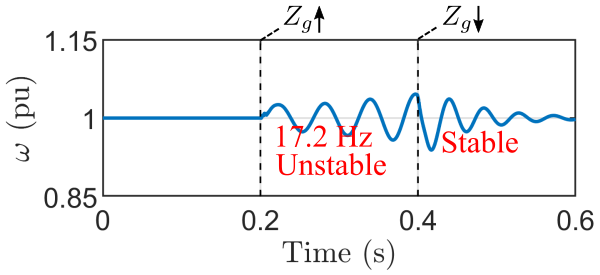}}
\caption{Illustration of synchronization stability with variation of grid impedance. 
(a) Root loci of $S_{i\theta}^\prime$ for grid-forming with $Z_g$ decreased from \highlight{$0.3(1/5+j)$~pu} to \highlight{$0.1(1/5+j)$~pu}. 
(b) Root loci of $S_{v\theta}^\prime$ for grid-following with $Z_g$ increased from \highlight{$0.4(1/5+j)$~pu} to \highlight{$0.6(1/5+j)$~pu}. (c) Simulation of grid-forming as $Z_g$ is reduced at \highlight{0.2~s} and changed back at \highlight{0.4~s}.
(d) Simulation of grid-following as $Z_g$ is increased at \highlight{0.2~s} and changed back at \highlight{0.4~s}.}
\label{Fig:RootLoci_GridStrength}
\end{figure}

\begin{figure}[t!]
\centering
	\subfloat[]{\includegraphics[scale = 0.75]{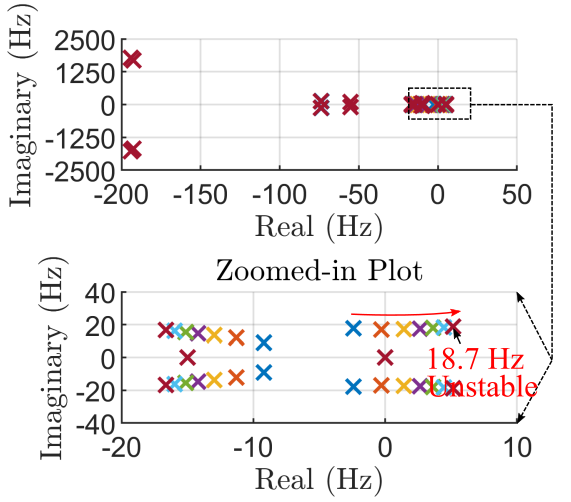}}
	\subfloat[]{\includegraphics[scale = 0.75]{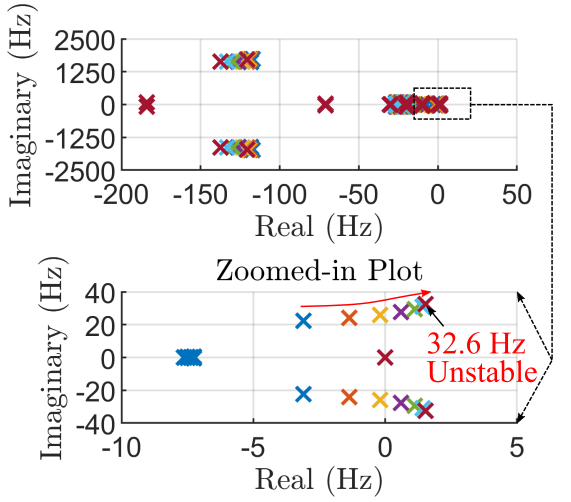}}
	
    \subfloat[]{\includegraphics[scale = 0.72]{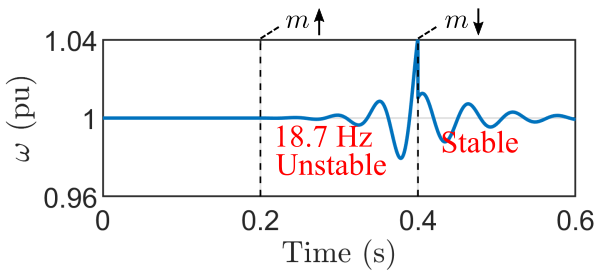}}
	\subfloat[]{\includegraphics[scale = 0.72]{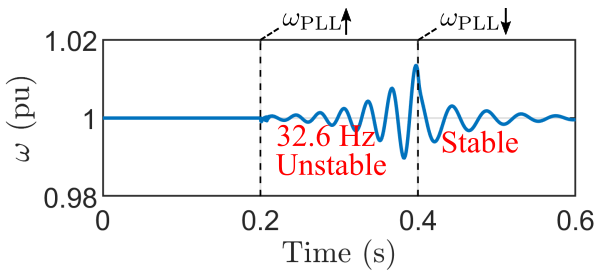}}
\caption{Illustration of synchronization stability with variation of synchronization controller gains. 
(a) Root loci of $S_{i\theta}^\prime$ for grid-forming with droop gain $m$ increased from \highlight{0.05~pu} to \highlight{0.2~pu}. 
(b)  Root loci of $S_{v\theta}^\prime$ for grid-following with PLL bandwidth $\omega_\text{PLL}$ increased from \highlight{15~Hz} to \highlight{60~Hz}.  
(c) Simulation of grid-forming as $m$ is increased at \highlight{0.2~s} and changed back at \highlight{0.4~s}; 
(d) Simulation of grid-following as $\omega_\text{PLL}$ is increased at \highlight{0.2~s} and changed back at \highlight{0.4~s}.}
\label{Fig:RootLoci_SynchroGain}
\end{figure}

\begin{figure}[t!]
\centering
	\subfloat[]{\includegraphics[scale = 0.75]{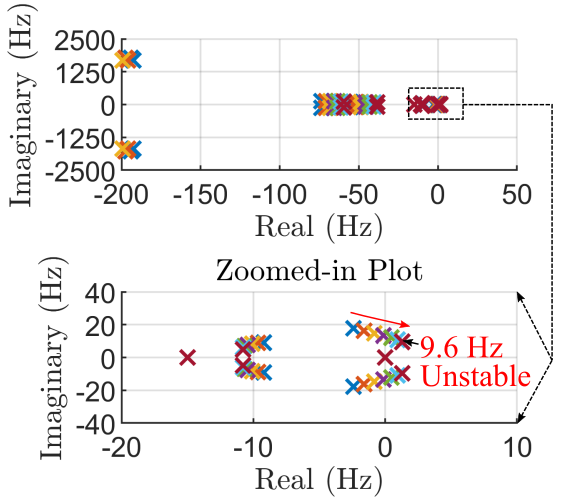}}
	\subfloat[]{\includegraphics[scale = 0.75]{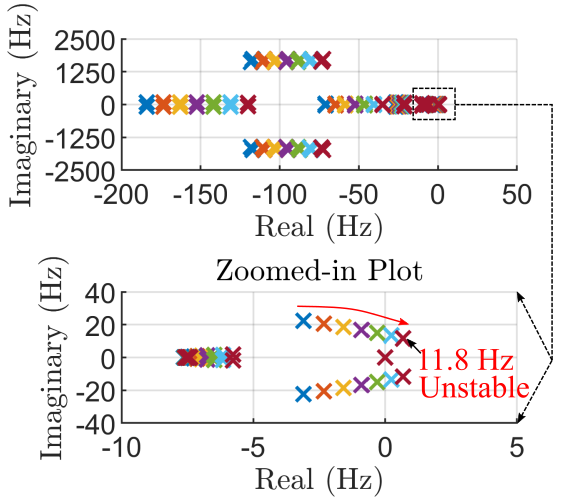}}
	
    \subfloat[]{\includegraphics[scale = 0.72]{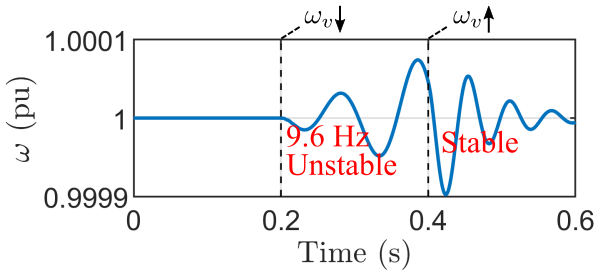}}
	\subfloat[]{\includegraphics[scale = 0.72]{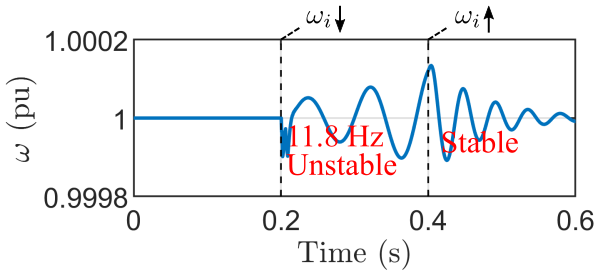}}
\caption{Illustration of synchronization stability with variation of inner-loop controller gains. 
(a) Root loci of $S_{i\theta}^\prime$ for grid-forming with voltage control bandwidth index $\omega_v$ reduced from \highlight{250~Hz} to \highlight{150~Hz}. 
(b) Root loci of $S_{v\theta}^\prime$ for grid-following with current control bandwidth index $\omega_i$ decreased from \highlight{250~Hz} to \highlight{150~Hz}.
(c) Simulation of grid-forming as $\omega_v$ is reduced at \highlight{0.2~s} and changed back at \highlight{0.4~s}. 
(d) Simulation of grid-following as $\omega_i$ is reduced at \highlight{0.2~s} and changed back at \highlight{0.4~s}.}
\label{Fig:RootLoci_InnerLoopGain}
\end{figure}


 To explore the factors that influence the stability of the synchronization of both forms of inverter, the root loci of $S_{i\theta}^\prime$ and $S_{v\theta}^\prime$ are plotted as three parameters are varied:  grid impedance in \figref{Fig:RootLoci_GridStrength}, synchronization gain in \figref{Fig:RootLoci_SynchroGain}, and inner-loop gain in \figref{Fig:RootLoci_InnerLoopGain}. In each case, sub-figures (a) and (c) are for frequency droop grid-forming and sub-figures (b) and (d) for PLL grid-following.
 
 \figref{Fig:RootLoci_GridStrength} shows that reducing $Z_g$ for the grid-forming inverter and reducing $Y_g$ for the grid-following inverter lead to the right-half plane roots and instability. In other words, a grid-following inverter is vulnerable to weak grid with low $Y_g$, which agrees with the consensus in the literature \citeref{wen2015analysis,dong2015analysis}. As a dual, a grid-forming inverter is vulnerable to strong grid with low $Z_g$. This observation is somewhat surprising at the first glance and has rarely been reported in literature \citeref{wang2020grid,rosso2021grid}, but follows from the duality that has been established. It also warns that the grid-forming inverter is not always a good choice. \textcolor{black}{Additionally, it is also worth mentioning that the grid-forming inverter investigated here has a double-loop control (inner current loop and outer voltage loop) for voltage forming and a droop control (active power frequency droop) for synchronization \cite{li2021impedance}. There are also alternative control options such as open-loop voltage control, single-loop voltage control, virtual synchronous generator, virtual impedance control, etc. Even though some control methods are proved to be equivalent \cite{li2021impedance,darco2014equivalence,meng2018generalized}, the stability characteristics may not be exactly same, which needs further investigations in the future.}
 
 \figref{Fig:RootLoci_SynchroGain} shows that increasing the droop gain $m$ of the grid-forming inverter (i.e., increasing $G_\text{FD}$ and $Z_\text{FD}$) and increasing the PLL bandwidth $\omega_\text{PLL}$ of the grid-following inverter (i.e., increasing $G_\text{PLL}$ and $Y_\text{PLL}$) lead to the instability. In other words, the larger of synchronization controller gains, the more prone inverters become to instability. 
 
 \figref{Fig:RootLoci_InnerLoopGain} shows that reducing the voltage loop bandwidth of the grid-forming inverter (i.e., increasing $Z_c$ \citeref{li2021impedance}) and reducing the current loop bandwidth of the grid-following inverter (i.e., increasing $Y_c$ \citeref{li2021impedance}) lead to the instability. In other words, the lower the inner-loop bandwidth, the greater the chance of unstable interaction. 
 
 Sub-figures (c) and (d) in \figref{Fig:RootLoci_GridStrength}, \figref{Fig:RootLoci_SynchroGain} and \figref{Fig:RootLoci_InnerLoopGain} show time-domain simulation results which confirm the theoretical analysis in (a) and (b). At \highlight{0.2 s}, parameters are tuned in the direction the root loci indicate which leads to the instability. The oscillations are observed to be at the frequencies predicted by root loci. At \highlight{0.4 s}, the parameters are returned to their initial values which re-stabilizes the system.

It is worth observing that the unstable modes discussed in this subsection are caused by insufficient damping coefficient (or damping torque) rather than insufficient synchronizing coefficient (or synchronizing torque). Insufficient damping coefficient occurs in a duality fashion: grid-forming with low grid impedance; grid-following with high grid impedance. This instability can generally able be avoided by tuning controllers (as seen in \figref{Fig:RootLoci_SynchroGain} and \figref{Fig:RootLoci_InnerLoopGain}) or by changing inner-loop control structures \citeref{li2021impedance}. Care is need, though, since there may be additional constraints on the controllers in practice which might limit the freedom to tune their parameters to enhance the small-signal stability. For example, the droop gain of grid-forming inverter is limited by the frequency variation limit specified in a grid code; and the inner loop bandwidths of inverters are limited by the switching and sampling frequency of the inverter hardware. In contrast to duality present in insufficient damping coefficient discussed above, insufficient synchronizing coefficient for both grid-forming and grid-following inverters occurs when the grid is very weak \citeref{kundur1994power,gu2021nature,dong2015analysis}. This case is related to the power transfer characteristic between two node voltages separated by an impedance. It is accompanied by the well-known limit on power transfer capability and is sensitive to the power flow \citeref{gu2021nature}. Under this condition there would be a direct loss of synchronization rather than an unstable oscillations. A further type of high-frequency instability is known and occurs when the bandwidth of the inner loop is too high. This is known as harmonic or super-synchronous instability \citeref{wang2018harmonic} and is distinctly different from the interaction instability caused by an inner loop that is too slow as in \figref{Fig:RootLoci_InnerLoopGain}.

 In conclusion, the imperative of stable synchronization of the two types of inverters places similar upper limits on synchronization controller gains and similar lower limits for the inner voltage- or current-loop controller gains but duality limits on grid strength, namely, a lower limit on the grid impedance for grid-forming but an upper limit on grid impedance (or low limit on admittance) for grid-following. All these observations point to the same conclusion: the larger the interaction terms in $S_{i\theta}^\prime$ and $S_{v\theta}^\prime$, the higher likelihood of instability. A summary of these discussions are also included in \tableref{Table:DualitySummary}.

%

\subsection{Rethinking Grid Strength}

\begin{figure}[t!]
\centering
\includegraphics[scale=1]{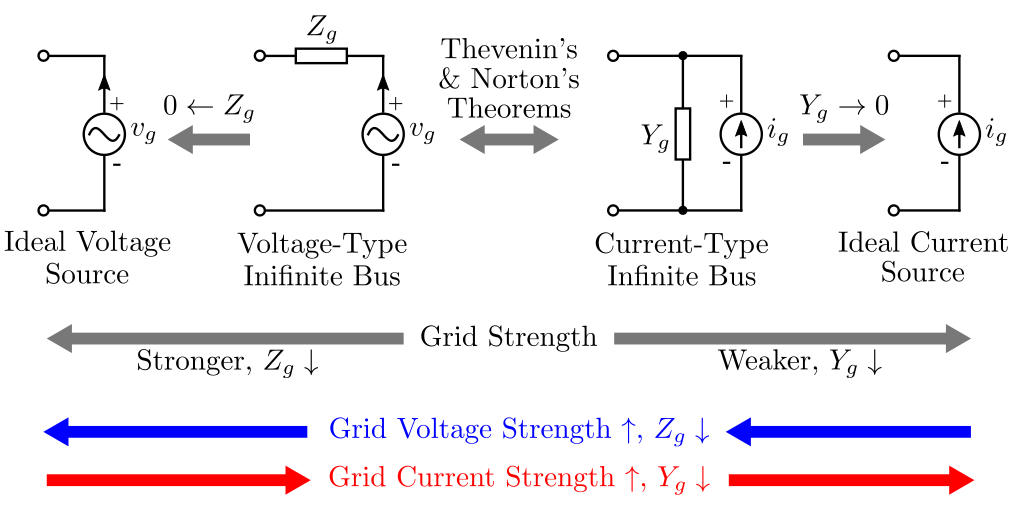}
\caption{Duality of grid strength.}
\label{Fig:GridStrength}
\end{figure}

As discussed in previous subsections, grid strength plays a role in inverter stability. In looking again at grid strength with the duality principle in mind, a normally overlooked fact is recalled that the term ``grid'' in grid strength is taken to mean a voltage grid rather than a current grid. Therefore, we might re-define the conventional grid strength more precisely as \textit{grid voltage strength} and, as a dual, also define the term \textit{grid current strength}, as illustrated in \figref{Fig:GridStrength}. Even though any source or bus can be represented as either voltage or current type through Thevenin's and Norton's theorems, the grid nowadays as we know is ``closer'' physically to a voltage source with relatively low equivalent series impedance. It has a high voltage strength when $Z_g$ is small and becomes an ideally strong (or stiff) voltage source when $Z_g = 0$. As a dual, a grid has a high current strength when $Y_g$ is small and becomes an ideally strong (or stiff) current source when $Y_g = 0$. A short circuit can be regarded as an ideal voltage source with $v_g = 0$ and $Z_g=0$; and open circuit can be regarded as an ideal current source with $i_g = 0$ and $Y_g=0$.

This wider interpretation of grid strength also helps to form intuitive understanding of the grid strength stability discussed in previous subsection. A grid-forming inverter (voltage-forming, current-following) is vulnerable to a strong grid and cannot operate with short-circuit (voltage-strong, current-weak). As a dual, a grid-following inverter (current-forming, voltage-following) is vulnerable to a weak grid and cannot operate with open-circuit (current-strong, voltage-weak). This is also summarized in \tableref{Table:DualitySummary}.



\subsection{A Multi-Inverter Power System} \label{Section:MultiInverter}

\begin{figure}[t!]
\centering
\includegraphics[scale=0.8]{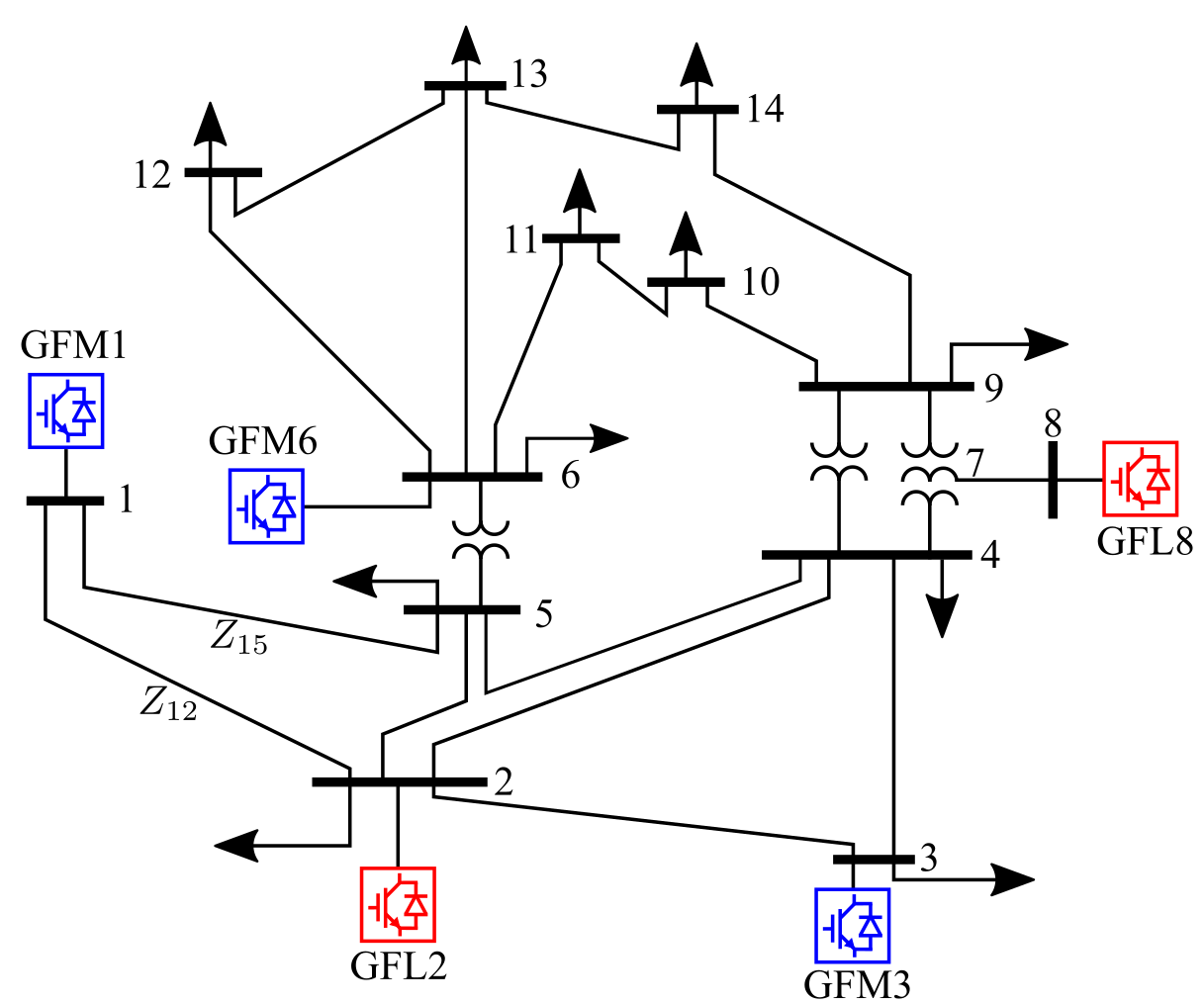}
\caption{A modified IEEE 14-bus inverter-based power system. (Notes: GFM indicates grid-forming, and GFL indicates grid-following).}
\label{Fig:14BusPowerSystem}
\end{figure}

\begin{figure*}[t!]
\centering
\includegraphics[scale=0.95]{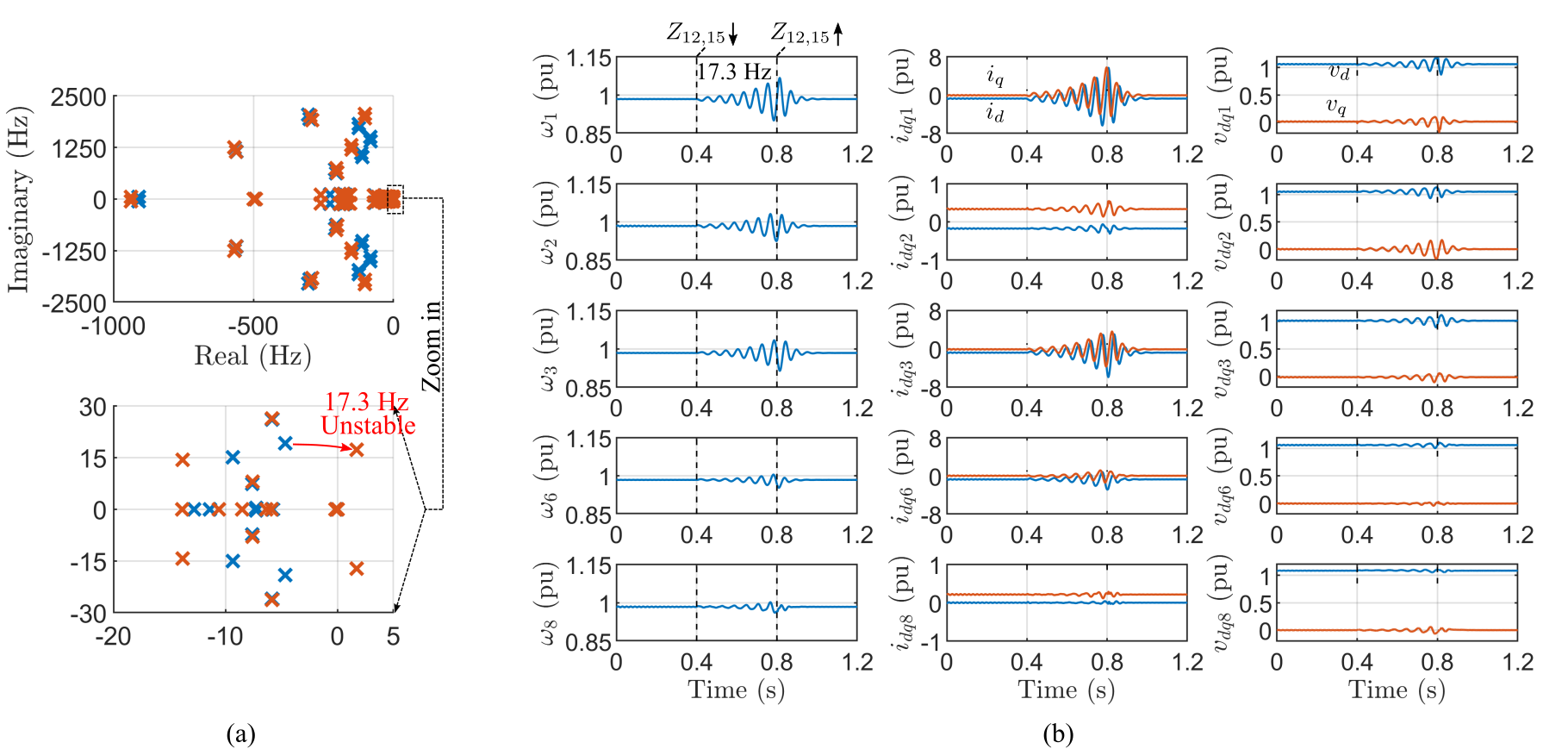}
\caption{(a) Pole loci of the whole-system with the standard line impedances (blue) and with $Z_{12}$ and $Z_{15}$ reduced to \highlight{one fifth} (red). (b) Simulation results for frequency, current and voltage at the five IBR buses. $Z_{12}$ and $Z_{15}$ are reduced at \highlight{0.4~s} and revert at \highlight{0.8~s}.}
\label{Fig:PoleLoci_Case4_GFM}
\end{figure*}

A network with several IBRs, shown in \figref{Fig:14BusPowerSystem} is tested next. It has the same layout and line impedances as the standard IEEE 14-bus test network \citeref{IEEE14Bus}, but all the synchronous generators have been replaced by either grid-forming inverters (blue) or grid-following inverters (red). The whole-system impedance model can be readily constructed by connecting the inverter impedance model in \figref{Fig:ModelDerivation} to the dynamic nodal admittance matrix of transmission lines \citeref{gu2021impedance,li2021mapping}. System stability can then be evaluated by observing the poles of the whole-system model in \figref{Fig:PoleLoci_Case4_GFM}(a). To study a change in grid strength, the impedances of lines $Z_{12}$ and $Z_{15}$, connecting GFM1, are reduced from the standard values of \highlight{$0.019+j0.059$ pu and $0.054+j0.22$ pu} (poles in blue) to \highlight{one fifth} of those values (poles in red).  This higher voltage strength connections leads to a \highlight{17.3 Hz} unstable oscillation (a strong-grid instability). This is confirmed by simulation results in \figref{Fig:PoleLoci_Case4_GFM}(b). At \highlight{0.4 s}, $Z_{12}$ and $Z_{15}$ are reduced, which leads to the unstable oscillation. At \highlight{0.8 s}, $Z_{12}$ and $Z_{15}$ are increased back to their standard values and the system becomes stable again.

As a dual, increasing the impedances of lines can also lead to the weak-grid instability of GFL2 or GFL8, which has been widely discussed in literature \citeref{wen2015analysis,dong2015analysis,yuan2017modeling,li2021mapping} and is not examined further here.

\section{Duality of Transient Stability} \label{Section:DualityTransientStability}

The analysis thus far (e.g., the characteristic equations in \figref{Fig:ModelDerivation} and the interaction analysis in \sectionref{Section:SingleInverterInfiniteBus}) has focused on small-signal instability, i.e., instability to perturbations around a given operating point. In practice, power grids are also required to be robust under abnormal operation (e.g., island operation) and severe disturbance (e.g., temporary short-circuit fault). This is known as the large-signal stability or transient stability. It is known that grid-forming inverters are established as a method for island operation with a formed grid voltage, duality can be used to operate an island with a formed grid current by grid-following inverters. Additionally, the transient stability of both grid-forming and -following inverters can be assessed in a similar manner to synchronous generators \citeref{kundur1994power} with the virtual inertia and virtual damping being considered in place of physical inertia and damping torque. These two aspects of duality will now be discussed.

\subsection{Single-Inverter System: Island Operation and Frequency Stability}
 \label{Section:SingleInverterPassiveLoad}

\begin{figure}[t!]
\centering
\includegraphics[scale=0.95]{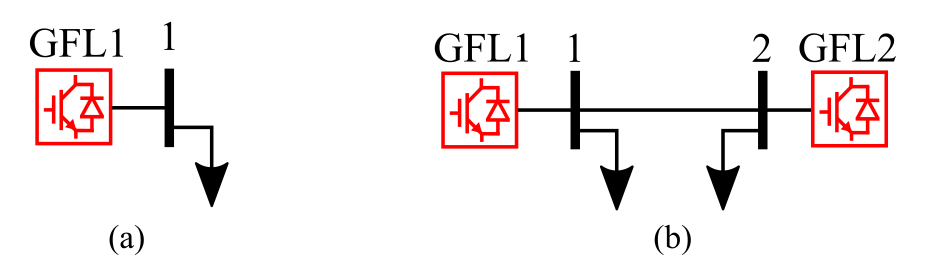}
\caption{System layout of the island operation of grid-following inverters. (a) Single-inverter case. (b) Two-inverter case.}
\label{Fig:IslandSystemGFL}
\end{figure}

\begin{figure}[t!]
\centering
\includegraphics[scale=0.95]{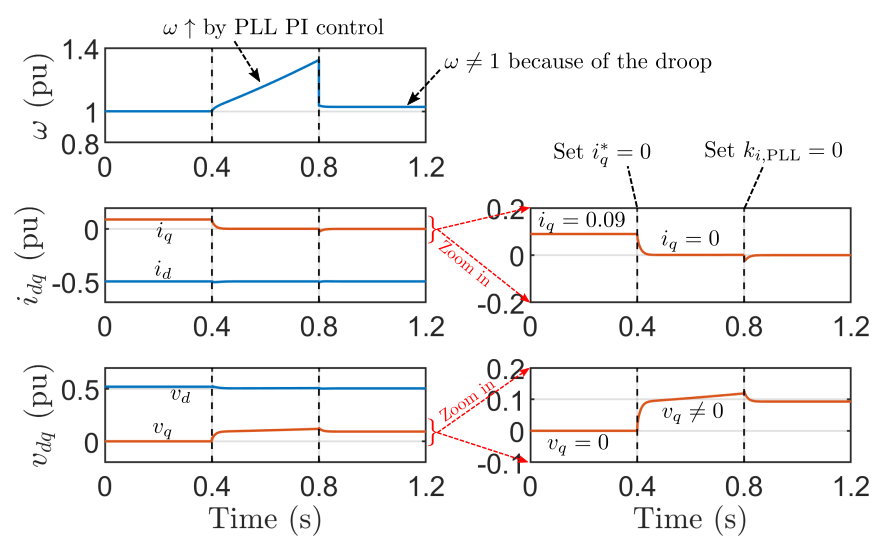}
\caption{Test results for the island operation of a grid-following inverter. $i_q^*$ is changed from \highlight{$0.09$ pu} to \highlight{0 pu} at \highlight{0.4~s} and $k_{i,\text{PLL}}$ is changed from its nominal value to 0 at \highlight{0.8 s}.}
\label{Fig:Test_PLLInverterPassiveLoad}
\end{figure}

The ability of grid-forming inverters to operate in electrical islands has been widely known for some time \citeref{li2021impedance,matevosyan2019grid}. In contrast, the ability of a grid-following inverter to do the same is controversial \citeref{dong2015analysis} because of the stereotype of ``following''. As discussed in \sectionref{Section:DualityGridFormingFollowing}, the grid-following inverter is more precisely a voltage-following current-forming inverter, and hence, it could form an island grid (current rather than voltage). To test this idea, a PLL grid-following inverter is simulated with a passive $RL$ load (\highlight{$1+j0.2$ pu}) in islanding mode. The system layout is shown in \figref{Fig:IslandSystemGFL} (a). The result in \figref{Fig:Test_PLLInverterPassiveLoad}, shows that from \highlight{0 s} to \highlight{0.4 s}, the inverter is stable and controls its current reference with \highlight{$i_d = i_d^* =  -0.5$ pu} and \highlight{$i_q = i_q^* = $ 0.09 pu}. Note that $v_d$ is approximately \highlight{0.5 pu} rather than \highlight{1 pu} because the voltage results from the inverter current and load impedance as determined by the Ohm's law. 

At \highlight{0.4 s}, $i_q^*$ is changed from \highlight{0.09} pu to \highlight{0 pu}, which, via Ohm's law, creates a non-zero $v_q$. The integrator of the PI controller in the PLL seeks to drive $v_q$ to zero by unremittingly increasing $\omega$. With the load and current references chosen, no equilibrium exists and the system frequency is unstable. 
At \highlight{0.8 s}, the integrator was disabled by setting $k_{i,\text{PLL}}=0$, which removes the contradiction and re-stabilizes the frequency. With the integrator removed, the PLL becomes a simple $v_q$-$\omega$ droop with a proportional droop gain of $k_{p,\text{PLL}}$. This further illustrates the duality between PLL and frequency droop and emphasises that some of the difference arise from presence or absence of integrator action.

In conclusion, the grid-following inverter is able to form grid current and operate in an islanding mode with a stable frequency but a widely varying grid voltage can occur because of the combination of the $i_{dq}^*$ setting and load impedance. Further, the PI function within the PLL can mean that there are current reference choices for which no equilibrium condition exists.

\color{black}

\subsection{Two-Inverter System: Angle Stability}

\begin{figure}[t!]
\centering
\includegraphics[scale=0.95]{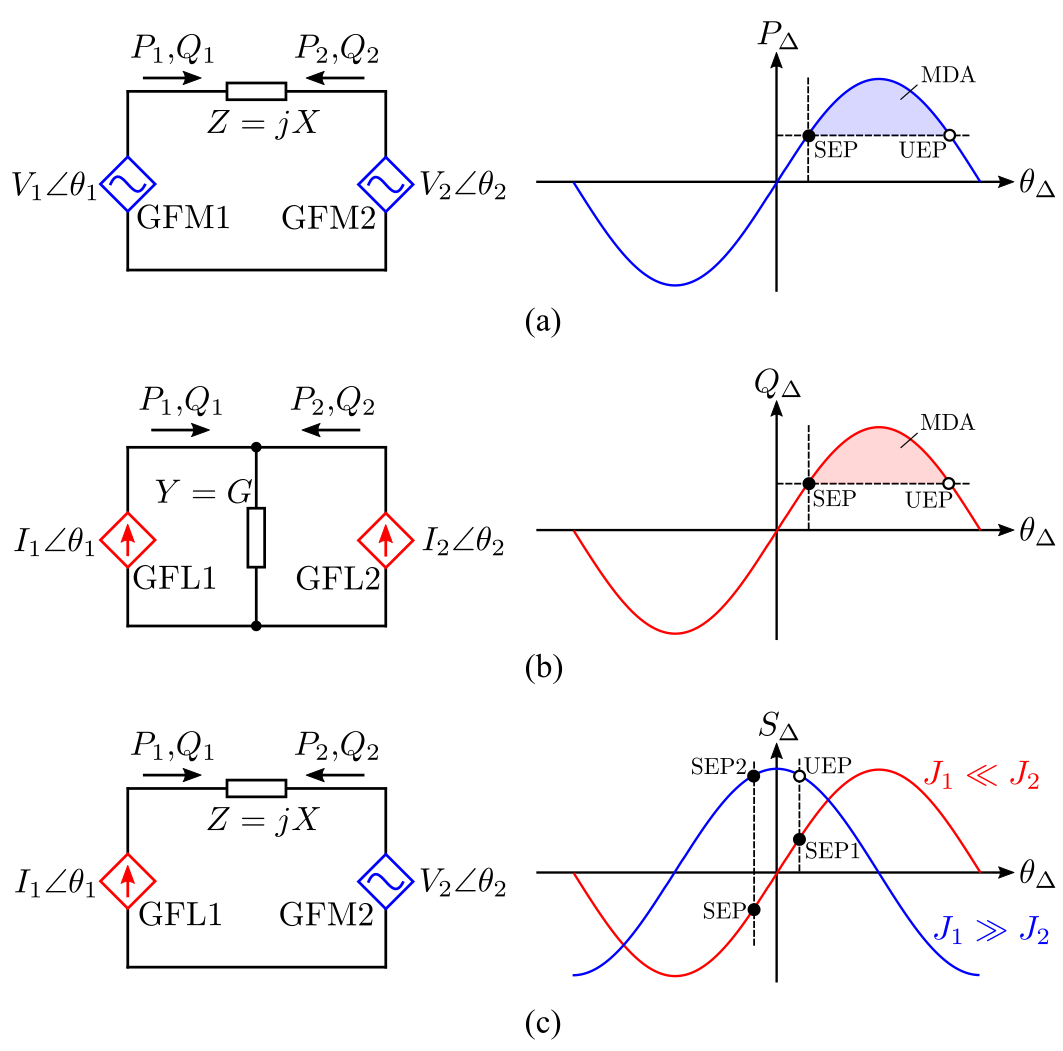}
\caption{\textcolor{black}{Angle stability analysis of different two-inverter systems. (a) Grid-forming inverters: Two-GFM system. (a) Grid-following inverters: Two-GFL system. (c) Interaction between grid-following and grid-forming inverters: GFM-GFL system.}}
\label{Fig:TwoInverterSystem}
\end{figure}

As analyzed in previous subsection, a single grid-following inverter can feed a load in island, and operate robustly, thanks to the duality between current-forming and voltage-forming and between PLL and frequency droop control. Here, we further explore the principle of the angle stability of different two-inverter systems in \figref{Fig:TwoInverterSystem}. 


Firstly, the system in \figref{Fig:TwoInverterSystem}(a) consisting of two grid-forming inverters (voltage sources) is discussed, whose transient stability is equivalent to the $P$-$\theta$ swing of a conventional two-synchronous-generator system \citeref{darco2014equivalence}. We briefly recall the principle again here. The large-signal swing interaction of two grid-forming inverters can be represented by
\begin{equation} \label{Equ:DeltaSwingGFM}
    J \ddot{\theta}_\Delta = P_\Delta^* - P_\Delta,
\end{equation}
where $J$ is the virtual inertia; $\theta_\Delta = \theta_1 - \theta_2$ is voltage angle difference; $P_\Delta^* = P_1^* - P_2^*$ is the difference of power setting points of droop control; $P_\Delta = P_1 - P_2$ is the difference of active powers. This swing interaction equation can be easily obtained by taking the difference of swing equations of two grid-forming inverters. When the impedance $Z$ is pure inductive $jX$ in \figref{Fig:TwoInverterSystem}(a), the synchronizing power $P_\Delta$ can be represented by
\begin{equation}
    P_\Delta = \frac{V_1V_2}{X} \sin(\theta_\Delta),
\end{equation}
i.e., sine $P_\Delta$-$\theta_\Delta$ relation. \figref{Fig:TwoInverterSystem}(a) plots this relation and indicates both a stable equilibrium point (SEP) and an unstable equilibrium point (UEP). The equilibrium is an SEP when $\theta_\Delta$ is less than $90\degree$. Further, the smaller of $\theta_\Delta$, the larger of the maximum decelerating area (MDA), the wider range of first-swing transient stability region \citeref{kundur1994power}.

\begin{figure}[t!]
\centering
\includegraphics[scale=0.7]{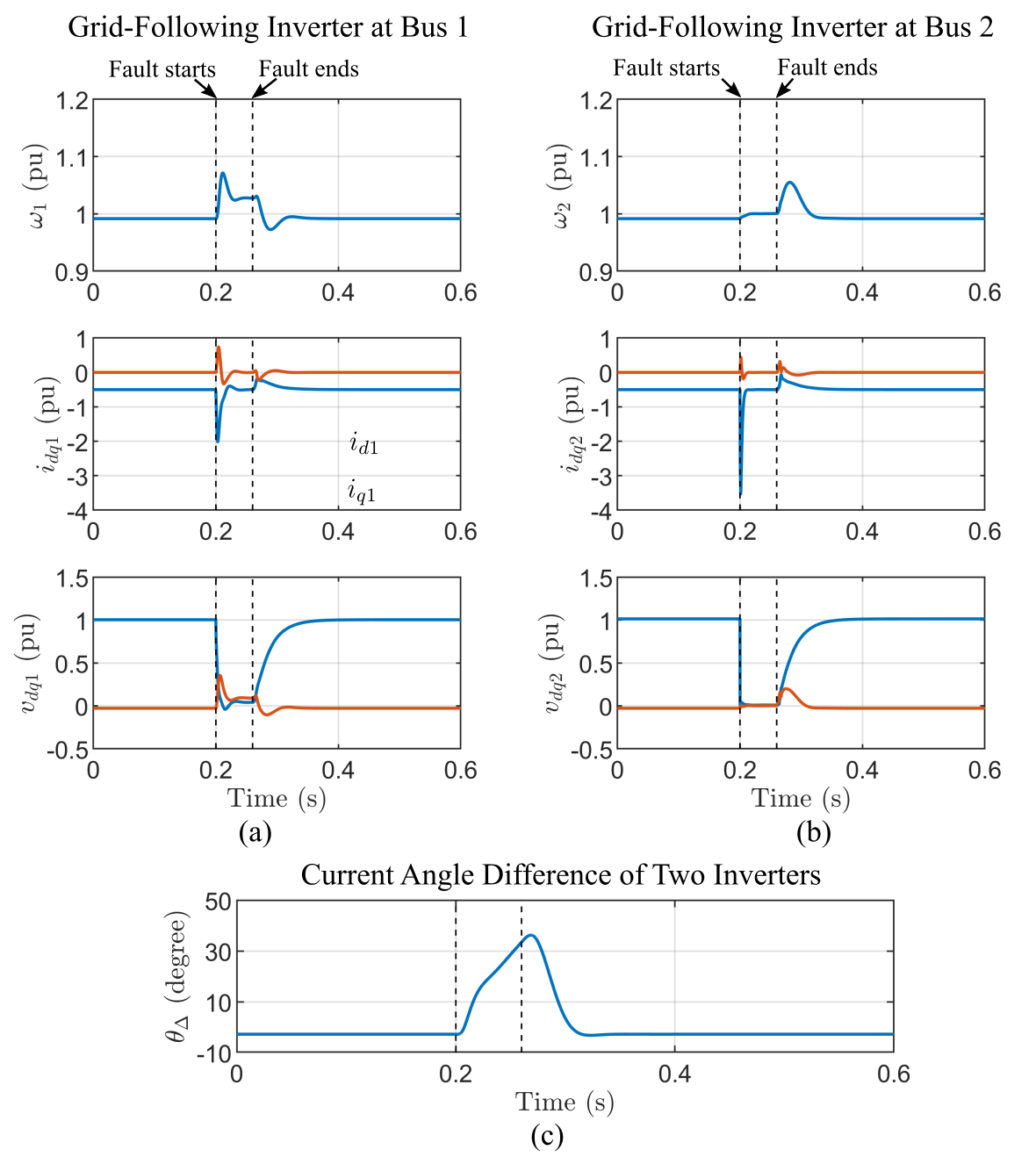}
\caption{Test results for the island operation of two grid-following inverters in \figref{Fig:IslandSystemGFL}(b). A temporary short-circuit fault happens to bus 2 at 0.2 s and is then cleared at 0.26 s. (a) GFL1. (b) GFL2. \textcolor{black}{(c) Current angle difference of GFL1 and GFL2.}}
\label{Fig:Sim_TwoGFL}
\end{figure}

Secondly, the system in \figref{Fig:TwoInverterSystem}(b) consisting of two grid-following inverters (current sources) is discussed. According to the analysis in \sectionref{Section:DualitySwing} and \sectionref{Section:SingleInverterIdealSource}, the grid-following inverters also have $V$-$\theta$ or $Q$-$\theta$ swing characteristics. We consider the reactive power format here,
\begin{equation} \label{Equ:DeltaSwingGFM}
    J \ddot{\theta}_\Delta = Q_\Delta^* - Q_\Delta.
\end{equation}
where $\theta_\Delta = \theta_1 - \theta_2$ is the current angle difference; $Q_\Delta^* = -(Q_1^* - Q_2^*)$ is the difference of the reactive power setting points used in PLL; $Q_\Delta = -(Q_1 -  Q_2)$ is the difference of reactive powers. When the admittance $Y$ in \figref{Fig:TwoInverterSystem}(b) is pure resistive $G$, we have
\begin{equation}
Q_\Delta = \frac{I_1 I_2}{G} \sin(\theta_\Delta).
\end{equation}
i.e., sine $Q_\Delta$-$\theta_\Delta$ relation. This means that the two-GFL system in \figref{Fig:TwoInverterSystem}(b) may have similar transient characteristics to the two-GFM system. In other words, two grid-following inverters should also be able to synchronize to each other without the existence of voltage sources (i.e., islanding mode) when the system impedance is resistance-dominated (i.e., passive loads connected). This is barely reported in literature and therefore is tested next. \figref{Fig:Sim_TwoGFL} shows the results. Before 0.2 s, both GFL1 and GFL2 operate stably. At 0.2 s, a temporary short-circuit fault occurs at bus 2. The fault reduces the voltage of two grid-following inverters and leads to surge current and a oscillation of frequency. After three fundamental periods, at 0.26 s, the temporary fault is cleared. GFL1 and GFL2 re-establish grid currents (through control of the inner loop), re-synchronize with each other (through control of the PLL or equivalently $v_q$-$\omega$ droop or equivalently $Q$-$\omega$ droop), and return the bus voltage back to its nominal value (through supply of current through the loads at buses).

\begin{figure}[t!]
\centering
\includegraphics[scale=0.75]{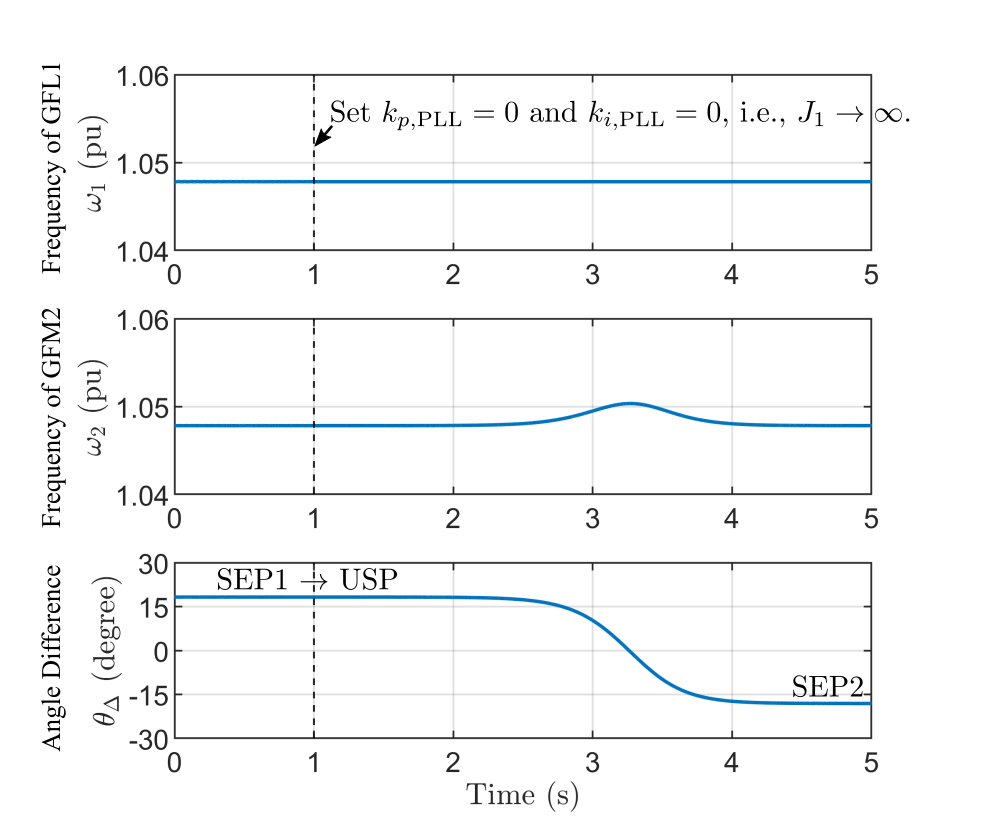}
\caption{\textcolor{black}{Test results of the synchronization between a grid-following and a grid-forming inverters in \figref{Fig:TwoInverterSystem}(c). At 1 s, the virtual inertia $J_1$ of GFL1 is increased to infinite by setting the PI controller gains of PLL to zero.}}
\label{Fig:Sim_GFL_GFM}
\end{figure}

Thirdly, the interaction between a grid-following and a grid-forming inverters is investigated, by studying the GFM-GFL system in \figref{Fig:TwoInverterSystem}(c). The swing equations of two inverters are
\begin{equation} \label{Equ:InteractonSwing}
\begin{aligned}
&
J_1 \ddot{\theta}_1 = -Q_1^* + \underbrace{XI_1^2 - V_2 I_1 \sin(\theta_1 - \theta_2)}_{Q_1}
\\
&
J_2 \ddot{\theta}_2 = P_2^* - \underbrace{(-V_2 I_1\cos(\theta_1 - \theta_2))}_{P_2},
\end{aligned}
\end{equation}
respectively. When $J_1 \gg J_2$, $\ddot{\theta}_1 \ll \ddot{\theta}_2$, $\ddot{\theta}_\Delta \approx -\ddot{\theta}_2$. This means the grid-forming inverter synchronizes to the grid-following inverter. Then, the whole system swing interaction equation is
\begin{equation}
J_2 \ddot{\theta}_\Delta = \underbrace{P_2^*}_{S_\Delta^*} - \underbrace{V_2 I_1\cos(\theta_\Delta)}_{S_\Delta},
\end{equation}
which is dominated by the second equation in \equref{Equ:InteractonSwing}, i.e., the swing dynamics of the grid-forming inverter GFM2. This equation indicates a cosine relation between the synchronizing power $S_\Delta$ and the angle difference $\theta_\Delta$, as plotted in blue in \figref{Fig:TwoInverterSystem}(c). By contrast, when $J_1 \ll J_2$, $\ddot{\theta}_1 \gg \ddot{\theta}_2$, $\ddot{\theta}_\Delta \approx \ddot{\theta}_1$. This means grid-following inverter synchronizes to the grid-forming inverter. The whole system swing interaction is dominated by the first equation in \equref{Equ:InteractonSwing}, i.e., the swing dynamics of the grid-following inverter GFL1, as
\begin{equation}
J_2 \ddot{\theta}_\Delta = \underbrace{-Q_1^*+XI_1^2}_{S_\Delta^*} - \underbrace{V_2 I_1\sin(\theta_\Delta)}_{S_\Delta}.
\end{equation}
This means a sine $S_\Delta$-$\theta_\Delta$ relation, as plotted in red in \figref{Fig:TwoInverterSystem}(c). It is worth mentioning that, for different values of $J_1/J_2$, the system has different power angle relation and different swing curves, and therefore different SEPs. This is also tested in \figref{Fig:Sim_GFL_GFM}. Before 1 s, the system operates stably with $J_1 \ll J_2$ by using the rated parameters in \tableref{Table:RatedParameters}. This corresponds to the SEP1 in \figref{Fig:TwoInverterSystem}(c) with $90\degree>\theta_\Delta>0\degree$. At 1 s, the virtual inertia $J_1$ of GFL1 is increased to infinite by setting the PI controller of PLL to zero, i.e., $k_{p,\text{PLL}}=0$ and $k_{i,\text{PLL}} = 0$. This sudden change leads to $J_1 \gg J_2$, and forces the system move from SEP1 ( red curve) to USP (blue curve) in \figref{Fig:TwoInverterSystem}(c). Then, the system gradually converges to SEP2 in \figref{Fig:TwoInverterSystem}(c) with $0\degree>\theta_\Delta>-90\degree$.

In conclusion, just as two grid-forming inverters that can synchronize with each other, two grid-following inverters can also synchronize with each other and operate stably. A grid-forming and grid-following inverters can also synchronize to each other depending on their inertia ratio. This further implies the duality between grid-forming and grid-following inverters. However, it is worth mentioning that, the specific power-angle ($S_\Delta$-$\theta$) relation and swing dynamics of the whole system are influenced by plenty of factors including the types of inverters (grid-forming, grid-following), virtual inertia values of inverters (large, small), impedance types of grid networks (resistive, inductive, capacitive, etc), and the steady-state operating points. This means that obtaining an interpretable and concise analytical solution of the transient swing dynamics of a multi-inverter system would be very challenging. In next subsection, the qualitative analysis of a multi-inverter system will be explored, with the help of time-domain simulations.

\color{black}

\subsection{Multi-Inverter System: Virtual Inertia and Virtual Damping}

In this subsection, parameters influencing the synchronization transient stability are further considered. The swing equations \equref{Equ:SwingGFM} and \equref{Equ:SwingGFL} derived in \sectionref{Section:SingleInverterIdealSource} are recalled here: For a grid-forming inverter, the virtual inertia $J$ and virtual damping coefficient $K_D$ are represented by $T_f/m$ and $m$ respectively, where $T_f$ is the time constant of the LPF and $m$ is the droop gain; For a grid-following inverter, the virtual inertia $J$ and virtual damping coefficient $K_D$ are represented by $1/k_{i,\text{PLL}}$ and $V_{d0}k_{p,\text{PLL}}/k_{i,\text{PLL}}$, where $k_{p,\text{PLL}}$ and $k_{i,\text{PLL}}$ are the gains the PLL PI controller. Note that $J$ and $K_D$ for the grid-forming inverter and $J$ for the grid-following inverter are not dependent on the steady-state operating point, which means they can be directly used for transient stability analysis. As for $K_D$ of the grid-following inverter, as long as $d$-axis voltage $V_{d0}$ is positive, $K_D$ is positively proportional to $k_{p,\text{PLL}}/k_{i,\text{PLL}}$. For both inverters, the smaller of $J$ and $K_D$, and the greater the chance of transient instability, like a synchronous generator. This principle will now be verified using the 14-bus system in \figref{Fig:14BusPowerSystem}.

\begin{figure}[t!]
\centering
\includegraphics[scale=0.7]{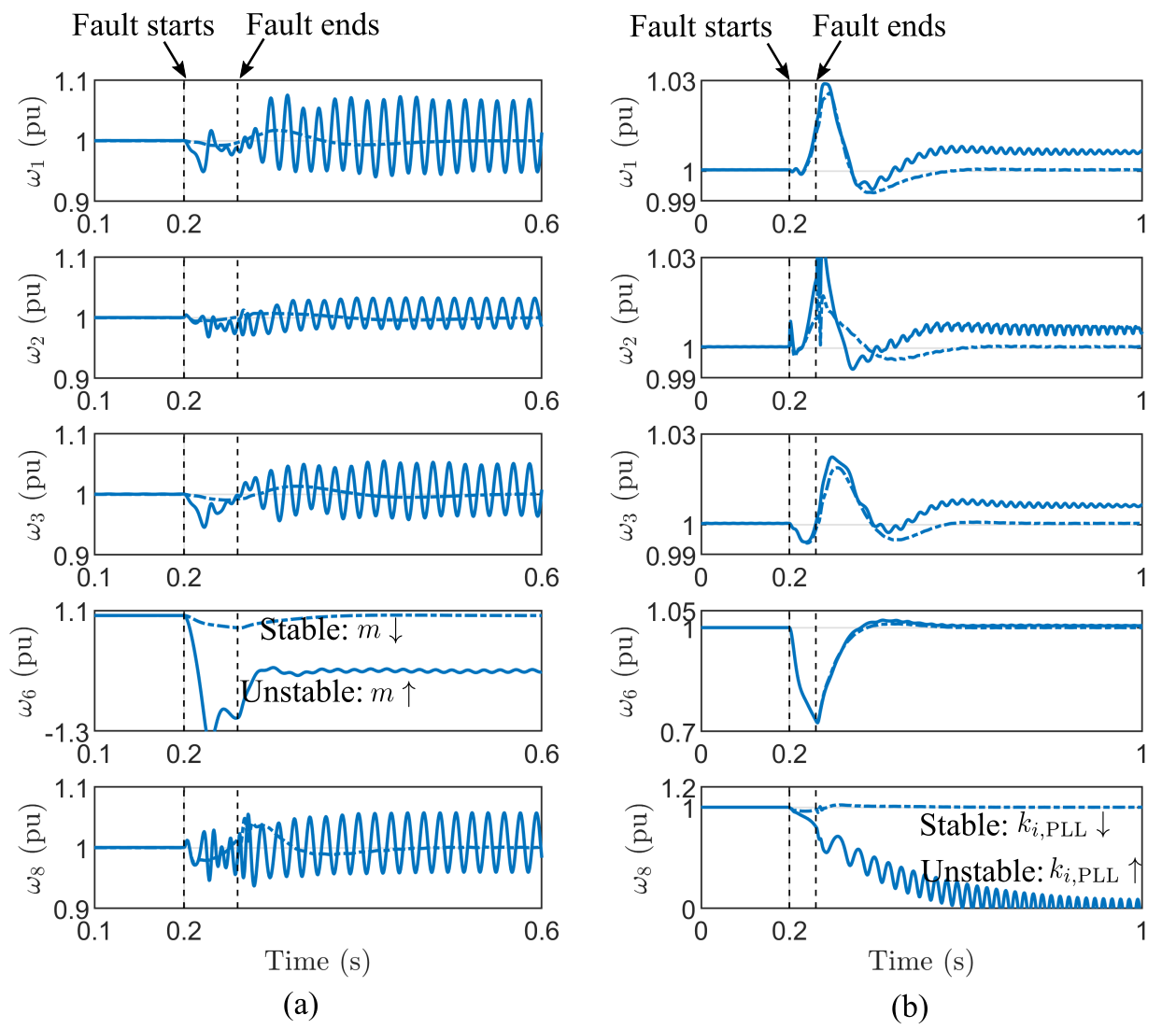}
\caption{Transient stability test when a temporary fault occurs in bus 6 at 0.2 s and is cleared at 0.26 s. The dashed line represents stable test and solid line indicates the unstable test. (a) Droop gain of grid-forming inverter is reduced to 0.01 pu (dashed line) or increased to 0.08 pu (solid line). (b) PLL integral gain of grid-following inverter is reduced by 10 times (dashed line) or increased by 10 times (solid line).}
\label{Fig:Test_TransientStability}
\end{figure}

Results from the test system are displayed in \figref{Fig:Test_TransientStability} with grid-forming examined in (a) and grid-following in (b). In both cases, a short-circuit fault was applied at 0.2 s at bus 6 (the approximate center in the grid). The fault was cleared three fundamental periods later \citeref{anderson1999power}, i.e., at 0.26 s. In \figref{Fig:Test_TransientStability}(a) results for different values of droop gain $m$ are shown. When the droop gain is small (0.01 pu), the system recovers to its pre-fault operating point after the fault is cleared (dashed line in the figure). When droop gain is large (0.08 pu, solid line in the figure), the system is seen to be small-signal stable before the fault occurs (before 0.2 s), but following the transient disturbance of the fault it does not recover to stable operation at its original operating point and is therefore has a large-signal instability. This observation is consistent with the swing equation \eqref{Equ:SwingGFM} for this case where a large droop gain creates a small inertia and small damping in the grid-forming inverter and as a result there is a greater chance of transient instability.

\figref{Fig:Test_TransientStability} (b) shows the transient stability with different values of the PLL integral gain $k_{i,\text{PLL}}$ of the grid-following inverter. When $k_{i,\text{PLL}}$ is small (dashed line), the system is stable. When $k_{i,\text{PLL}}$ is large (solid line), the system is small-signal stable, but does not recover to its original operating point after the fault event and therefore is large-signal unstable. This observation also consistent with the swing equation \equref{Equ:SwingGFL} because a large $k_{i,\text{PLL}}$ indicates small virtual inertia and small damping of the grid-following inverter. 

Comparing \figref{Fig:Test_TransientStability}(a) and (b), it is seen that the grid-forming inverter GFM6 suffers a loss of synchronism and frequency collapse in (a) whereas it is the grid-following inverter GFL8 that looses synchronism in (b). In both cases, but particularly there are disturbances to other IBR but no loss of synchronism. This also illustrates that the main source of the transient instability is different in grid-forming and grid following cases.


\section{Conclusions} \label{Section:Conclusions}

An examination of synchronization loops of grid-forming and grid-following inverters has revealed a set of properties which are duals of each other. This extends well beyond the simple observation that one acts as a voltage source and the other as a current source. The duality reveals similarities and differences between the two types that can aid our understanding of how instability can arise in inverter networks and what measures can be taken to combat instability. The duality is seen in the following points.

a) Converse synchronization loop. The $P$-$\omega$ droop control for a grid-forming inverter is equivalent to $i_d$-$\omega$ droop control when $v_d$ is constant and $v_q$ is 0, which can further be regarded as a PLL acting on $i_d$ with a proportional phase-locking controller. The PLL for a grid-following inverter acting on $v_q$ can be regarded as $v_q$-$\omega$ droop control with a PI droop gain and further is equivalent to $Q$-$\omega$ droop control when $i_d$ is constant and $i_q$ is 0.

b) Converse grid-interfacing characteristics. The grid-forming inverter is current-following and voltage-forming. The grid-following inverter is voltage-following and current-forming.

c) Converse swing characteristics. The grid-forming inverter has a current-angle swing or, equivalently, a active-power-angle swing. The grid-following inverter has a voltage-angle swing or, equivalently, reactive-power-angle swing.

d) Similar controller gain characteristics (converse impedance/admittance shaping). For grid-forming and grid-following, the larger the synchronization controller gains (frequency droop gain and PLL bandwidth) or the smaller the inner-loop controller gains (voltage- and current-loop bandwidths), the higher the likelihood of interaction and synchronization instability.

e) Converse grid strength compatibility. The grid-forming inverter is vulnerable to strong grid with low grid impedance (strong grid voltage and weak grid current). The grid-following inverter is vulnerable to weak grid with low grid admittance (strong grid current and weak grid voltage).

f) Similar transient stability mechanism. The transient stability of both the grid-forming inverter and grid-following inverter depends on their swing equation, virtual inertia, and virtual damping. Like a grid-forming inverter, a grid-following inverter can also operate in island mode or synchronize with another grid-following inverter without a voltage source present.

The duality interpretation proposed here has been shown to unify the view of grid-synchronization and grid-interfacing characteristics of two types of inverter in a symmetric and elegant form. This unified view can not only offer promising pathway towards multi-inverter analysis which is missing now \citeref{gu2021nature}, but also inspire novel concepts (such as $i_d$-PLL, $v_q$-$\omega$ droop, voltage/current-forming/following, and voltage/current strength) and help to interpret surprising phenomena (such as strong grid instability of grid-forming inverters and islanding of grid-following inverters). It is anticipated that the use of duality can also be applied to or incorporated with other power engineering theories and lead to further new interpretations of future power systems. But it is also remarked that not all properties are duals because some proprieties relate to the voltage-dominated sources, parallel-connected apparatuses, and inductor-dominated lines of grids nowadays, and the presence or absence of integral action in the synchronization loop. This may lead to different roles of two types of inverters in different research areas (voltage stability, power quality, etc), which needs further studies.



\appendices

\section{Fundamentals of Complex $dq$ Frame} \label{Appendix:ComplexFrame}

The complex $dq$ frame ($dq\pm$ frame) \citeref{li2021impedance,li2020interpretating,harnefors2007modeling} is used for small-signal analysis in this article. For an arbitrary signal $u_{dq}$ in $dq$ frame, the corresponding complex signal $u_{dq\pm}$ in $dq\pm$ frame can be obtained as
\begin{equation}
\underbrace{\begin{bmatrix} u_+ \\ u_- \end{bmatrix}}_{u_{dq\pm}}
=
\underbrace{\begin{bmatrix} 1 & j \\ 1 & -j \end{bmatrix}}_{T_{j}}
\underbrace{\begin{bmatrix} u_d \\ u_q \end{bmatrix}}_{u_{dq}}
\end{equation}
where $u_+ = u_d + ju_q$ and $u_- = u_d - ju_q$ are the forward and backward complex space vectors, respectively; and $T_j$ is the transformation matrix. According to this transformation law, for a $dq$ frame system model
\begin{equation}
\underbrace{
\begin{bmatrix} y_d \\ y_q \end{bmatrix}
}_{y_{dq}}
=
\underbrace{
\begin{bmatrix} G_{dd} & G_{dq} \\ G_{qd} & G_{qq} \end{bmatrix}
}_{G_{dq}}
\underbrace{
\begin{bmatrix} u_d \\ u_q \end{bmatrix}
}_{u_{dq}}
\end{equation}
we can get the corresponding $dq\pm$ frame model as
\begin{equation}
\underbrace{
\begin{bmatrix} y_+ \\ y_- \end{bmatrix}
}_{y_{dq\pm}} 
= 
\underbrace{
\begin{bmatrix} G_+ & G_- \\ \overline{G}_- & \overline{G}_+ \end{bmatrix}
}_{G_{dq\pm}} 
\underbrace{
\begin{bmatrix} u_+ \\ u_- \end{bmatrix}
}_{u_{dq\pm}}
\end{equation}
with
\begin{equation}
G_{dq\pm} = T_j G_{dq} T_j^{-1}
\end{equation}
where the overbar in $G_{dq\pm}$ indicates the conjugate operation. For power system applications, $G_{dq\pm}$ normally has a simpler and more symmetric mathematical structure than $G_{dq}$, which facilitates the analysis \citeref{li2021impedance,li2020interpretating}. 

\section{Parameters} \label{Section:Paramters}

If not specified, rated parameters in \tableref{Table:RatedParameters} are used when obtaining the results of single-inverter cases in \sectionref{Section:DualitySmallSignalStability}. For the  14-bus multi-inverter case, parameters are available online at \citeref{FuturePowerNetworks}.

\begin{table}[H]
\renewcommand{\arraystretch}{1.3}
\newcommand{\tabincell}[2]{\begin{tabular}{@{}#1@{}}#2\end{tabular}}
\caption{Rated Parameters.}
\label{Table:RatedParameters}
\centering
\begin{tabular}{l | c | c }
\hline
\hline
\multicolumn{3}{c}{Frequency Droop Grid-Forming Inverter}
\\
\hline
Frequency Droop Bandwidth & $\omega_f$ & 15 Hz
\\
\hline
Frequency Droop LPF & LPF & $\frac{1}{1+s/\omega_f}$
\\
\hline
Frequency Droop Gain & $m$ & 0.05 pu
\\
\hline
Voltage Control Bandwidth Index & $\omega_v$ & 250 Hz
\\
\hline
\hline
\multicolumn{3}{c}{PLL Grid-Following Inverter}
\\
\hline
PLL Bandwidth & $\omega_\text{PLL}$ & 15 Hz
\\
\hline
PLL Proportional Controller & $k_{p,\text{PLL}}$ & $\omega_\text{PLL}$
\\
\hline
PLL Integral Controller & $k_{i,\text{PLL}}$ & $\omega_\text{PLL}^2/4$
\\
\hline
Current Control Bandwidth Index & $\omega_i$ & 250 Hz
\\
\hline
\hline
\multicolumn{3}{c}{AC Filter}
\\
\hline
$LC$ & \tabincell{c}{$\Omega_0 L$\\$\Omega_0 C$} & \tabincell{c}{0.05 pu\\0.02 pu}
\\
\hline
\hline
\multicolumn{3}{c}{Grid}
\\
\hline
Base/Fundamental Frequency & $\Omega_0$ & 50 Hz
\\
\hline
\hline
\end{tabular}
\end{table}









\ifCLASSOPTIONcaptionsoff
  \newpage
\fi

\bibliographystyle{IEEEtran}
\bibliography{Paper_Syn}

\end{document}